\documentclass[twocolumn,showpacs,preprintnumbers,amsmath,amssymb,prb]{revtex4-1}
\usepackage{amsfonts,amsmath,amssymb}
\usepackage{bbm}
\usepackage[english]{babel}
\usepackage[utf8]{inputenc}
\usepackage[T1]{fontenc}
\usepackage{times}
\usepackage{mathrsfs}
\usepackage{graphicx}
\usepackage{dcolumn}
\usepackage{bm}
\usepackage[colorlinks,bookmarks=true,citecolor=blue,linkcolor=red,urlcolor=blue]{hyperref}
\usepackage[tight, FIGTOPCAP, normal, raggedright, nooneline]{subfigure}
\usepackage{hyperref}

\renewcommand{\vec}[1]{\boldsymbol{\mathbf{#1}}}
\renewcommand{\vr}{\vec r}
\newcommand{\vk}{\vec k}

\newcommand{\vPW}{\vec K_0}
\newcommand{\PW}{K_0}
\newcommand{\ve}{\vec e}
\newcommand{\vq}{\vec q}

\newcommand{\vnabla}{\mbox{\boldmath $\nabla$}}

\begin{document}

\title{Josephson effect in a Weyl SNS junction}
\author{Kevin A.~Madsen$^1$}
\author{Emil J.~Bergholtz$^2$}
\author{Piet W.~Brouwer$^1$}
\affiliation{$^1$Dahlem Center for Complex Quantum Systems and Institut f\"ur Theoretische Physik, Freie Universit\"at Berlin, Arnimallee 14, 14195 Berlin, Germany \\
$^2$Stockholm University, Department of Physics, SE-106 91 Stockholm, Sweden}

\date{\today}

\begin{abstract}
We calculate the Josephson current density $j(\phi)$ for a Weyl superconductor--normal-metal--superconductor junction for which the outer terminals are superconducting Weyl metals and the normal layer is a Weyl (semi)metal. We describe the Weyl (semi)metal using a simple model with two Weyl points. The model has broken time-reversal symmetry, but inversion symmetry is present. We calculate the Josephson current for both zero and finite temperature for the two pairing mechanisms inside the superconductors that have been proposed in the literature, zero-momentum BCS-like pairing and finite-momentum FFLO-like pairing, and assuming the short-junction limit. For both pairing types we find that the current is proportional to the normal-state junction conductivity, with a proportionality coefficient that shows quantitative differences between the two pairing mechanisms. The current for the BCS-like pairing is found to be independent of the chemical potential, whereas the current for the FFLO-like pairing is not.
\end{abstract}

\maketitle
\section{Introduction}\label{intro}

A Weyl semimetal, a semimetal in which conduction and valence bands have nondegenerate touching points, is a paradigm of gapless topological matter.\cite{weylturnerreview,weylhosurreview} Weyl semimetals were initially proposed theoretically, but meanwhile they have been observed experimentally in a range of materials. \cite{weylexp1,weylexp2,weylexp3,weylexp4,weylexp5,weylexp6,weylexp7,weylexp8,weylexp9,weylexp10,weylexp11,weylexp12,weylexp13,weylexp14,weylexp15,weylexp16,weylexp17,weylexp18}  Electrons in a Weyl semimetal with momentum and energy in the vicinity of the band touching points are described by an effective low-energy theory which has the same form as the Weyl Hamiltonian of massless relativistic particles. 

Already before their experimental discovery, the question of possible forms of superconductivity in Weyl semimetals was considered theoretically.\cite{weylsc1,weylsc2,weylsc3,weylsc4,weylsc5,Meng_PRB_86_054504,Lu_PRL_114_096804} Weak attractive interactions were found to lead to unconventional superconducting states in Weyl (semi)metals, even if the chemical potential is not at the nodal point. (That situation is sometimes referred to as a ``Weyl metal.'') Examples of superconducting phases that were predicted are Fulde-Ferrell-Larkin-Ovchinnikov (FFLO)-like states in which the Cooper pairs have a finite momentum, or a more standard Bardeen-Schrieffer-Cooper (BCS)-like state in which the Cooper pairs have zero momentum, but the spectrum proves to have nodal points. Recently, the experimental observation of a superconducting phase has been reported for the compounds MoTe$_2$\cite{weylscexp} and TaP,\cite{arXiv_1611_02548} which are believed to be a Weyl metal in the normal state, although the nature of the observed superconducting phase is not known yet.

In this article we theoretically investigate the Josephson effect in a Weyl superconductor--normal-metal--superconductor (SNS) junction in which the superconducting terminals are such superconducting Weyl metals and the normal spacer layer is a Weyl (semi)metal, too. A possible realization of such a structure could consist of three layers of essentially the same Weyl semimetal, but with different doping levels, such that the more strongly doped outer layers have entered the superconducting phase, whereas the weakly doped central layer remains normal. For a simple but generic model of two Weyl cones\cite{weylsc4} --- the minimal model for a Weyl semimetal, since Weyl cones always have to come in pairs --- we calculate the temperature dependence of the Josephson current and the current-phase relationship for both types of pairing. Our work complements studies by Kanna {\em et al.}, who also consider the Josephson effect for an SNS junction in which the normal spacer is a Weyl semimetal, but with conventional BCS superconductors for the outer terminals,\cite{josephsonweylcompare} and by Kim {\em et al.}, who study four-terminal junctions with mixed Weyl superconductor and conventional superconductor contacts.\cite{PRB_93_214511}

A Weyl semimetal can be considered a three-dimensional generalization of graphene, a system for which the Josephson effect has received abundant experimental attention.\cite{Heersche_Nature_446_56_2007,Heersche_Solid_State_Commun_143_72_2007,Kanda_Physica_C_470_1477_2010,English_PRB_94_115435_2016,Kurter_Nat_Commun_6_7130_2015,Komatsu_PRB_86_115412_2012,Shailos_EPL_79_57008_2007,Du_PRB_77_184507_2008,Mizuno_Nat_Commun_4_2716_2013} Indeed, we find that the theory of the Josephson effect in a Weyl (semi)metal-based Josephson junction closely resembles the theory for the Josephson effect in graphene,\cite{josephsongraphene1,beenakker_RMP_80_1337_2008} but only for the case of FFLO-like pairing, and is also similar to the theory of Josephson junctions on the surface of three-dimensional topological insulators. \cite{Olund2012, Ghaemi2016} The case of BCS-like pairing is different, however, in the sense that the normalized critical Josephson current is effectively independent of the chemical potential.

The remainder of this article is organized as follows: Our model will be explained in detail in Sec.\ \ref{model}, closely following Refs.\ \onlinecite{weylsc4} and \onlinecite{andreevweyl1}. We calculate the Josephson current using the scattering approach, which is explained in Sec.\ \ref{scattering}. A summary of the results is given in Sec.\ \ref{results} and we conclude in Sec.\ \ref{conclusions}. The appendix contains details of the calculation.

\section{Model}\label{model}

\begin{figure}[h!]
	\centering
	\includegraphics[trim=0 -2.5cm 0 0, width=.23\textwidth]{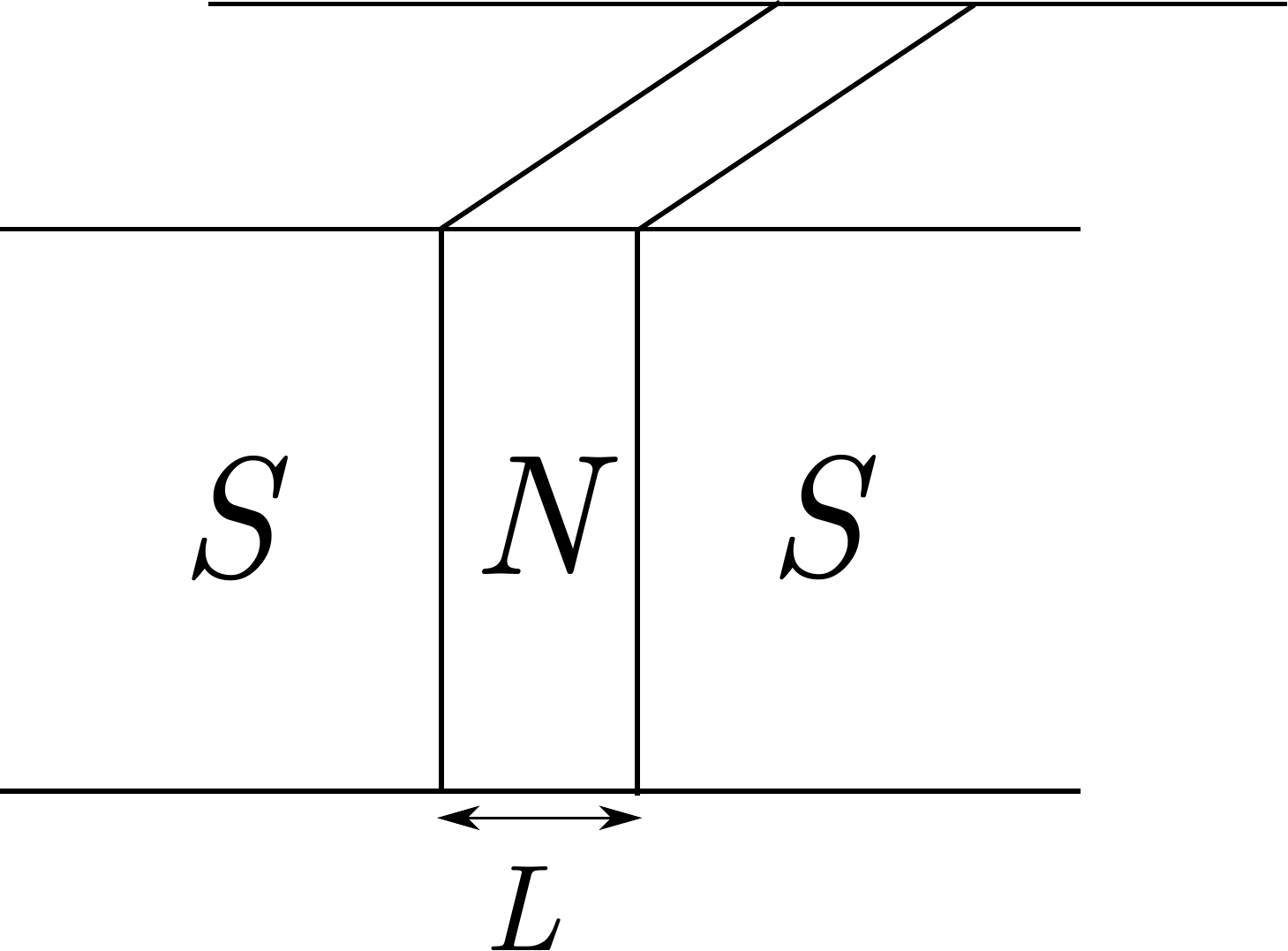}
	\includegraphics[width=.23\textwidth]{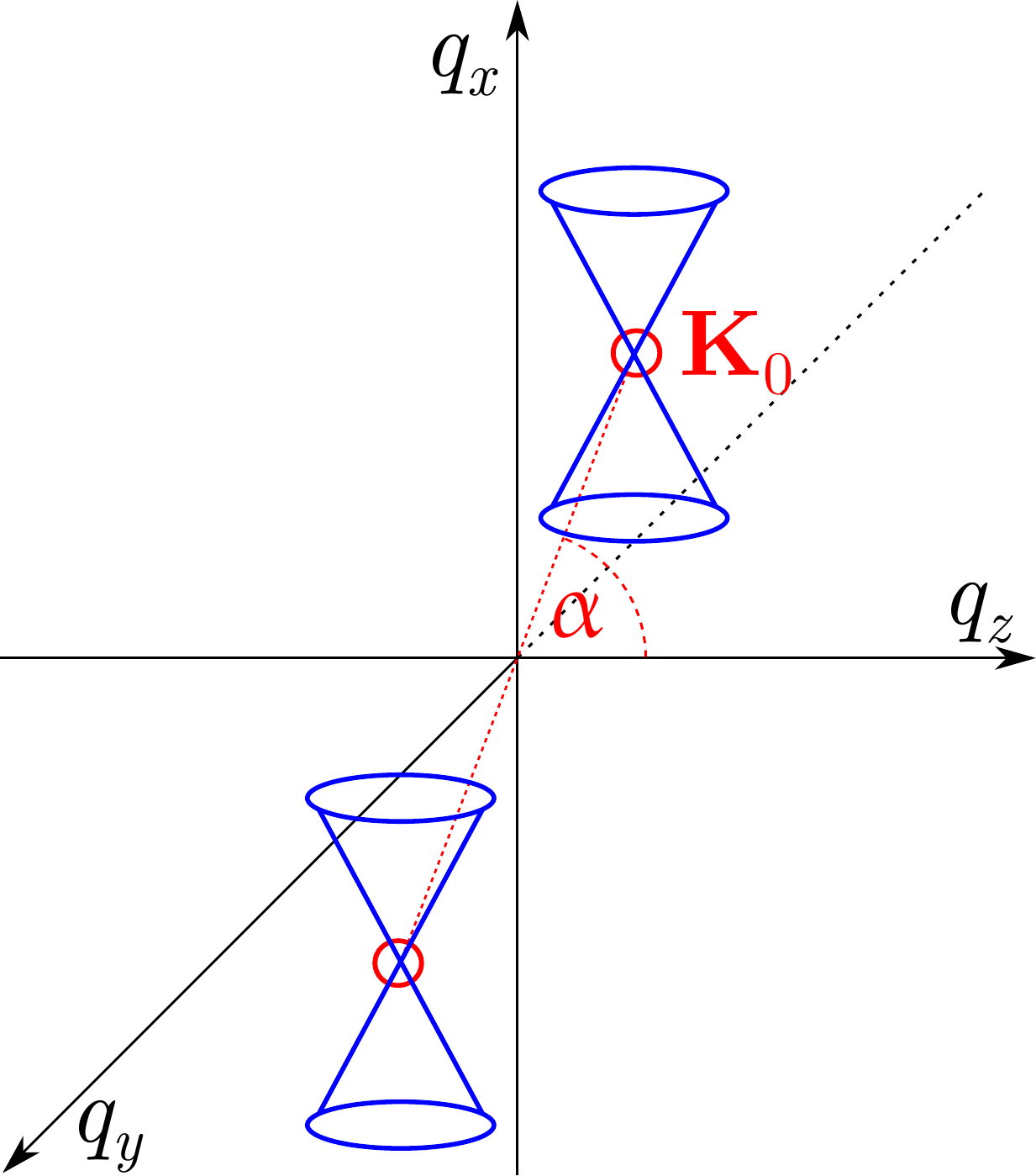}
	\caption{(Left) Sketch of the geometry under consideration: An intrinsic Weyl (semi)metal for $-L/2 < z < L/2$ is sandwiched between doped, superconducting Weyl metals for $z < -L/2$ and $z > L/2$. (Right) Schematic of the momentum space with two ``Weyl cones'' at momenta $\pm \vPW$. The angle between $\vPW$ and the $z$ axis is denoted $\alpha$.}
	\label{fig:layout}
\end{figure}

We consider a slab of normal Weyl (semi)metal with two Weyl points of opposite chirality at momenta $\pm \vPW$ for $-L/2 < z < L/2$, sandwiched between two superconducting regions, which are modeled as heavily doped Weyl semimetals --- also called ``Weyl metals'' --- with superconducting order; see Fig.\ \ref{fig:layout}. Without loss of generality we may assume that the positions of the Weyl points are in the $p_x$--$p_z$ plane, $\vPW = \PW(\ve_z \cos \alpha + \ve_x \sin \alpha)$ and that the Weyl point at $\vPW$ has positive chirality. Reference \onlinecite{weylsc4} has proposed a two-band Hamiltonian that describes this situation, and we also adopt this Hamiltonian for our present calculations. The normal-state Hamiltonian $h_{\pm}$ for momenta $\vk = \pm \vPW + \vq$ in the vicinity of $\pm \vPW$ reads \cite{weylsc4}
\begin{align}
  h_{\pm}(\vq) =&\, \hbar v [q_1 \sigma_1 + q_2 \sigma_2 \mp q_3 \sigma_3]
  \nonumber \\ & \mbox{} + V_0 \theta(|z| - L/2) - \mu,
  \label{eq:Ham}
\end{align}
where $v$ is the Fermi velocity, $\mu$ the chemical potential,
\begin{align}
  q_1 &= q_x \cos \alpha - q_z \sin \alpha, \nonumber \\
  q_2 &= q_y, \\
  q_3 &= q_z \cos \alpha + q_x \sin \alpha, \nonumber
\end{align}
and, similarly, $\sigma_1 = \sigma_x \cos \alpha - \sigma_z \sin \alpha$, $\sigma_2 = \sigma_y$, $\sigma_z = \sigma_z \cos \alpha + \sigma_x \sin \alpha$. The potential $V_0$ is a large negative potential offset realizing the high carrier density in the superconducting regions at $z < -L/2$ and $z > L/2$.\cite{andreevweyl1} The Hamiltonian (\ref{eq:Ham}) has broken time-reversal symmetry. However, it has inversion symmetry, $h_{-}(\vq) = \sigma_z h_+(-\vq) \sigma_z$, which ensures that both Weyl points are at the same energy.

In Weyl semimetals there are two possible superconducting pairings.\cite{weylsc1,weylsc2,weylsc3,weylsc4,weylsc5,andreevweyl1} For the BCS-like pairing the Cooper pairs have zero momentum and zero spin; they combine electrons from Weyl points with opposite chirality. For the FFLO-like pairing the Cooper pairs have zero spin and finite momentum $\pm 2 \vPW$, and consist of two electrons of the same chirality. The FFLO-like pairing is energetically favorable for a lattice model with weak attractive interactions underlying the continuum Hamiltonian (\ref{eq:Ham}),\cite{weylsc4,weylsc5} although other models point to the BCS-like paired state as the superconducting ground state,\cite{weylsc1,weylsc3} so that we must consider both pairing forms here. If both time-reversal symmetry and inversion symmetry are broken, the degeneracy between the two Weyl points may be lifted, and the FFLO-like pairing is the only pairing form allowed by symmetry.\cite{weylsc1}

The Bogoliubov-de Gennes Hamiltonian for four-component Nambu vectors $[\psi_{\uparrow}(\vr),\psi_{\downarrow}(\vr),\psi^{\dagger}_{\downarrow}(\vr),-\psi^{\dagger}_{\uparrow}(\vr)]$ reads\cite{andreevweyl1}
\begin{align}
	H_{\rm B} &= 	\left(
	\begin{array}{cc}
		h_{+}(-i \vnabla - \vPW)  & \Delta(z) \\
		\Delta(z)^* & - h_{-}(-i \vnabla - \vPW) \\
	\end{array}
	\right), \label{eq_bcsdbdg}
\end{align}
for the BCS-like pairing, where we used that $\vq = -i \vnabla \mp \vPW$ at the Weyl point $\pm \vPW$, and that $\sigma_y h_-^*(-i \vnabla + \vPW) \sigma_y = h_-(-i \vnabla - \vPW)$. In this case a doubling of the Hilbert space is not necessary, since the BCS-like pairing involves electrons at different Weyl points. On the other hand, the FFLO-like pairing involves electrons at the same Weyl node, so that doubling of the Hilbert space is necessary and there are separate Bogoliugov-de Gennes Hamiltonians $H_{{\rm F},\pm}$ for each Weyl node,
\begin{align}
  H_{{\rm F},\pm} &=
	\left(
	\begin{array}{cc}
		h_{\pm} (-i \vnabla \mp \vPW) & \Delta(z) e^{\pm 2 i \vPW \cdot \vr} \\
		\Delta(z)^* e^{\mp 2 i \vPW \cdot \vr} & - h_{\pm} (-i \vnabla \pm \vPW) \\
	\end{array}
	\right). \label{eq_fflodbdg}
\end{align}
The superconducting order parameter $\Delta(z)$ is zero for $-L/2 < z < L/2$, and it is set to $\Delta(z) = \Delta_0 e^{- i \phi/2}$ for $z < - L/2$ and $\Delta(z) = \Delta_0 e^{i \phi/2}$ for $z > L/2$, corresponding to a difference $\phi$ between the phases of the superconducting order parameters. We consider the short junction limit $L \ll \xi = \hbar v/\Delta_0$, which allows us to neglect the energy dependence of the momenta in the normal region $-L/2 < z < L/2$. The potential offset $V_0$ for the superconducting regions is chosen to be large and negative, but still small enough that the linearized description around each Weyl cone continues to apply.

For FFLO-like pairing the excitation spectrum of the superconductor is gapped, with excitation gap,
\begin{equation}
  \Delta_{\rm F} = \Delta_0.
\end{equation}
For BCS-like pairing the quasiparticle spectrum in the two superconducting regions has two nodes, turning the superconducting regions in to ``Weyl superconductors''.\cite{weylsc4,weylsc1,weylsc3,weylsc5} In the limit that the potential offset $V_0$ is large and negative, effectively only momenta along the $z$ axis need to be considered, for which the excitation gap is\cite{weylsc4,weylsc5}
\begin{equation}
  \Delta_{\rm B} = \Delta_0 |\sin \alpha|;
\end{equation}
see also the appendix.

\section{Scattering approach}\label{scattering}

\begin{figure}
\centering
	\includegraphics[width=.3\textwidth]{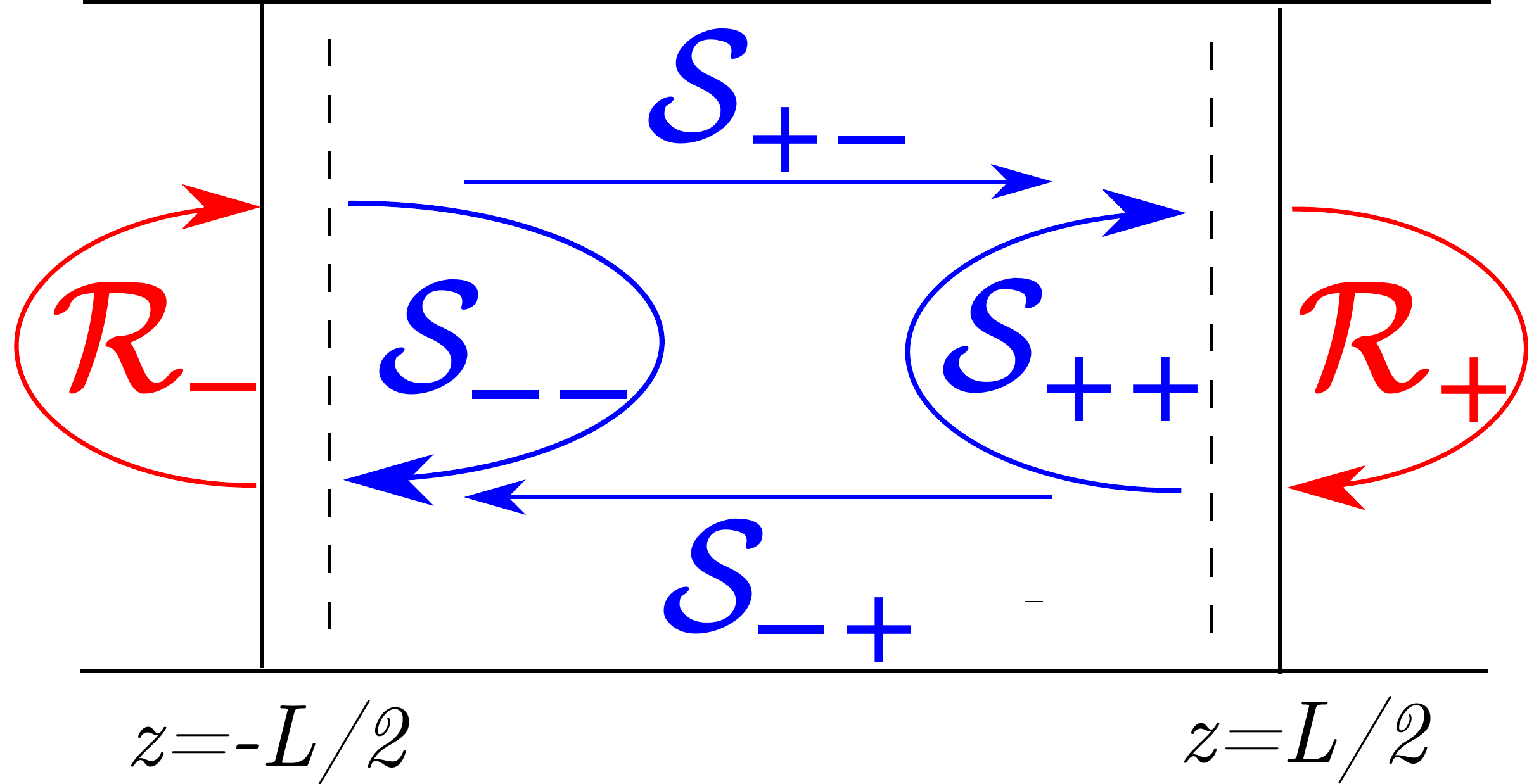}
	\caption{\label{fig:scattering} Definition of the scattering matrices ${\cal R}_+$, ${\cal R}_{-}$, ${\cal S}_{++}$, ${\cal S}_{+-}$, ${\cal S}_{-+}$ and ${\cal S}_{--}$ for a Josephson junction.}
\end{figure}

The Josephson current and the Andreev bound state energies are calculated following the approach of Refs.\ \onlinecite{beenakker_PRL67_3836_1991,beenakker_in_Transport_Phenomena_in_Mesoscopic_Systems_eds_H_Fukuyama_ad_T_Ando_Springer_Berlin_1992,josephsontemperature}, which describe the normal spacer layer and the superconducting interface in terms of their reflection and transmission matrices. For the normal spacer layer these are the reflection matrices ${\cal S}_{--}$ and ${\cal S}_{++}$ for reflection of excitations incident from the negative and positive $z$ direction, respectively, as well as the transmission matrices ${\cal S}_{-+}$ and ${\cal S}_{+-}$; see Fig.\ \ref{fig:scattering}. For the superconducting interfaces one needs the reflection matrices ${\cal R}_{\pm}$ that describe normal and Andreev reflection at the interface at $z = \pm L/2$. Since the wave numbers $q_x$ and $q_y$ parallel to the interface are conserved, all reflection and transmission matrices are diagonal in $q_x$ and $q_y$. They do, however, retain a $2 \times 2$ matrix structure to accommodate for the particle and hole degrees of freedom. The reflection and transmission matrices of the normal spacer layer are diagonal with respect to the particle and hole degrees of freedom; the reflection matrices ${\cal R}_{\pm}$ mix particle and hole degrees of freedom.

For the Josephson current density $j(\phi)$ one then finds\cite{josephsontemperature}
\begin{align}
  j(\phi) =& - \frac{4 e k_{\rm B} T}{\hbar}
  \frac{d}{d \phi}
  \int \frac{d q_x dq_y}{(2 \pi)^2}
  \sum_{m=0}^{\infty}
  \ln \det A(q_x,q_y,\phi;i \omega_m),
\end{align}
where
\begin{align}
  A = \left( \begin{array}{cc} \openone & 0 \\ 0 & \openone \end{array} \right)
  - \left( \begin{array}{cc}
  {\cal R}_- & 0 \\ 0 & {\cal R}_+ \end{array} \right)
  \left( \begin{array}{cc} {\cal S}_{--} & {\cal S}_{-+} \\
  {\cal S}_{+-} & {\cal S}_{++} \end{array} \right),
  \label{eq:A}
\end{align}
where $\openone$ is the $2 \times 2$ matrix unit matrix in particle-hole space and $i \omega_m = i (2 m + 1) \pi k_{\rm B} T$ the (imaginary) energy the scattering matrices have to be evaluated at. The dependence on the superconducting phase difference $\phi$ enters through the reflection matrices ${\cal R}_{\pm}$ for the two superconducting interfaces. In the short-junction limit we may neglect the dependence of the normal-spacer scattering matrix ${\cal S}$ on the energy $i \omega_m$. The energies of the Andreev bound states can be found from the same matrix $A$, from the condition $\det A(q_x,q_y,\phi;\varepsilon) = 0$, with real energy $|\varepsilon| < \Delta_0$.

The calculation of the scattering matrices ${\cal R}$ and ${\cal S}$ appearing in Eq.\ (\ref{eq:A}) proceeds by wave function matching and is described in detail in the appendix. We here list the results for $\det A$ for the cases of BCS-like and FFLO-like pairing,
\begin{align}
  \det A_{\rm F} =&\, 4 e^{2 i \gamma_{\rm F}}
  \left[\frac{\varepsilon^2}{\Delta_{\rm F}^2} - 1 +
  \frac{q_z^2 \sin^2(\phi/2)}{q_z^2 + q_{\perp}^2 \sin^2(q_z L)} \right], 
  \label{eq:detAF} \\
  \det A_{\rm B} =&\, 4 e^{2 i \gamma_{\rm B}}
  \left[\frac{\varepsilon^2}{\Delta_{\rm B}^2}
  - \sin^2\varphi 
  \right. \nonumber \\ &\, \left. \mbox{}
  + \frac{q_z^2(\sin^2(\phi/2) - \cos^2 \varphi)}{q_z^2 + q_{\perp}^2 \sin^2(q_z L)} \right],
  \label{eq:detAB}
\end{align}
where $q_{\perp}^2 = q_x^2 + q_y^2$, $q_z^2 = (\mu/\hbar v)^2 - q_{\perp}^2$, $\tan \varphi = q_y/q_x$, and
\begin{equation}
  \gamma_{\rm F,B} = -\arccos \frac{i \omega_m}{\Delta_{\rm F,B}}.
\end{equation}
For the Andreev bound states the condition $\det A(q_x,q_y,\phi;\varepsilon) = 0$ then gives
\begin{align}
  \varepsilon^2 = 
  \Delta_{\rm F}^2 \left[1 - \frac{q_z^2 \sin^2(\phi/2)}
  {q_z^2 + q_{\perp}^2 \sin^2(q_z L)} \right]
  \label{eq:epsF}
\end{align}
for the case of FFLO-like pairing and
\begin{equation}
  \varepsilon^2 =
  \Delta_{\rm B}^2 
  \left[\sin^2 \varphi - \frac{q_z^2(\sin^2(\phi/2) - \cos^2 \varphi)}
  {q_z^2 + q_{\perp}^2 \sin^2(q_z L)}\right]
  \label{eq:epsB}
\end{equation}
for BCS-like pairing.
The zero-temperature Josephson current can also be calculated from the $\phi$ dependence of the Andreev bound state energy,
\begin{equation}
  j(\phi) = - \frac{2e}{\hbar}
  \int \frac{d q_x dq_y}{(2 \pi)^2} \frac{\partial \varepsilon}{\partial \phi},
\end{equation}
where $\varepsilon$ is the positive root of Eq.\ (\ref{eq:epsF}) or (\ref{eq:epsB}).

\section{Results}\label{results}

We present results for the Josephson current density normalized by the normal-state conductivity $\sigma_{\rm N}$ of the junction. The normal-state conductivity, defined as the conductance per unit cross-sectional area, is calculated from the Landauer formula (see Ref.\ \onlinecite{rngraphene} for the corresponding result in two dimensions),
\begin{equation}
  \sigma_{\rm N} = \frac{2 e^2}{h} \int \frac{dq_x dq_y}{(2 \pi)^2}
  T(q_x,q_y) \label{eq:GN}
\end{equation}
where the transmission probability
\begin{equation}
  T(q_x,q_y) = \frac{q_z^2}{q_{z}^2 + q_{\perp}^2 \sin^2(q_z L)},
  \label{eq:TN}
\end{equation}
as shown in the appendix. 

For normal transport through the junction, two regimes may be distinguished: transport that is dominated by propagating modes or by evanescent modes. In the former case one has $|\mu| L/\hbar v \gg 1$ and the integral in Eq.\ (\ref{eq:GN}) is dominated by $q_{\perp} < |\mu|/\hbar v$, corresponding to real $q_z$. The junction conductivity is
\begin{equation}
  \sigma_{\rm N} = \frac{\mu^2 e^2}{6 \pi^2 \hbar^3 v^2}.
  \label{eq:sigmaN1}
\end{equation}
In the latter case the integral in Eq.\ (\ref{eq:GN}) is dominated by $q_{\perp} > |\mu|/\hbar v$, which corresponds to imaginary $q_z$, and the junction conductivity is\cite{Baireuther_PRB_89_035410,Sbierski_PRL_113_026602}
\begin{equation}
  \sigma_{\rm N} = \frac{e^2 \ln 2}{2 \pi^2 \hbar L^2}.
  \label{eq:sigmaN2}
\end{equation}

\begin{figure}[h!]
	\centering
	\includegraphics[width=.23\textwidth]{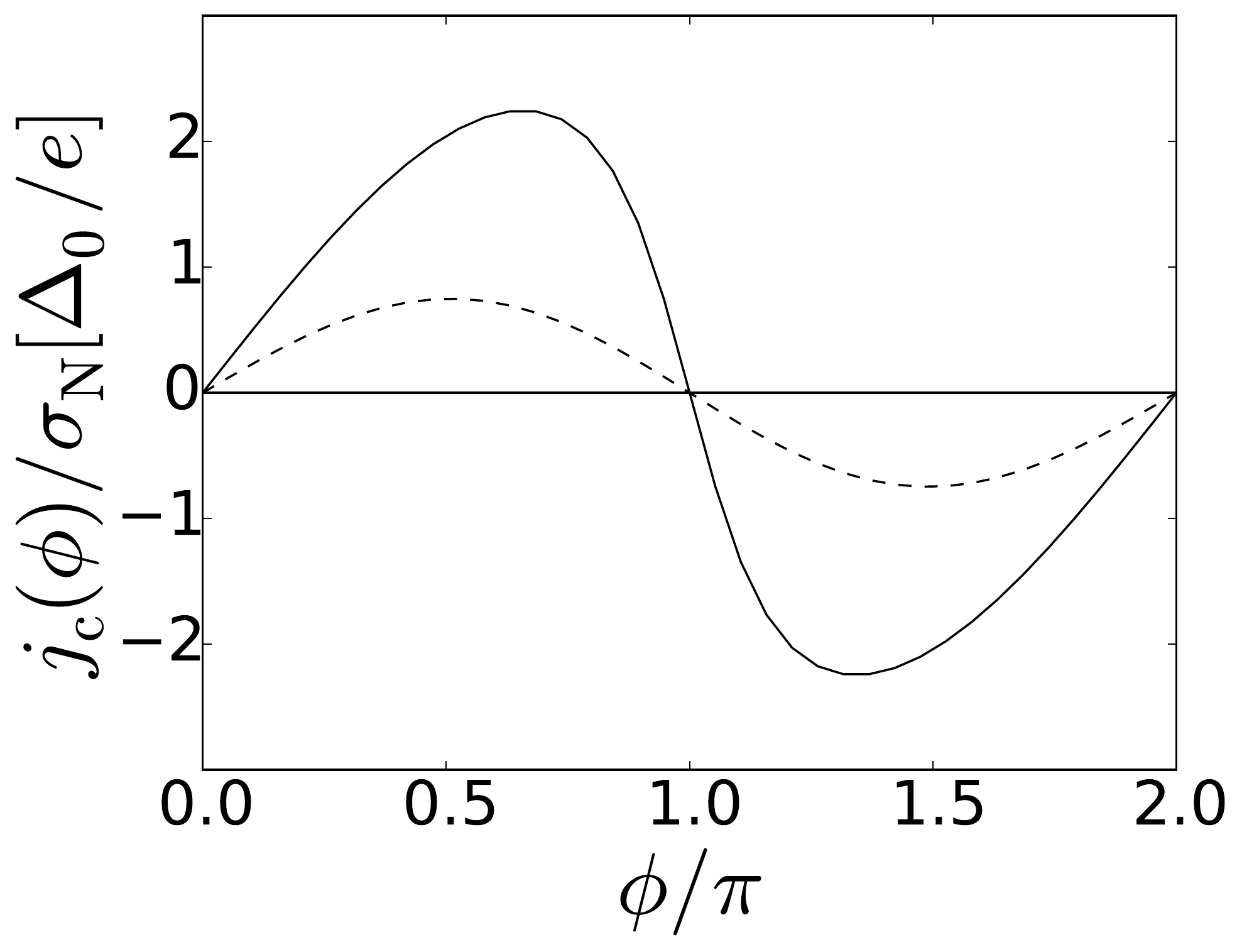}
	\includegraphics[width=.23\textwidth]{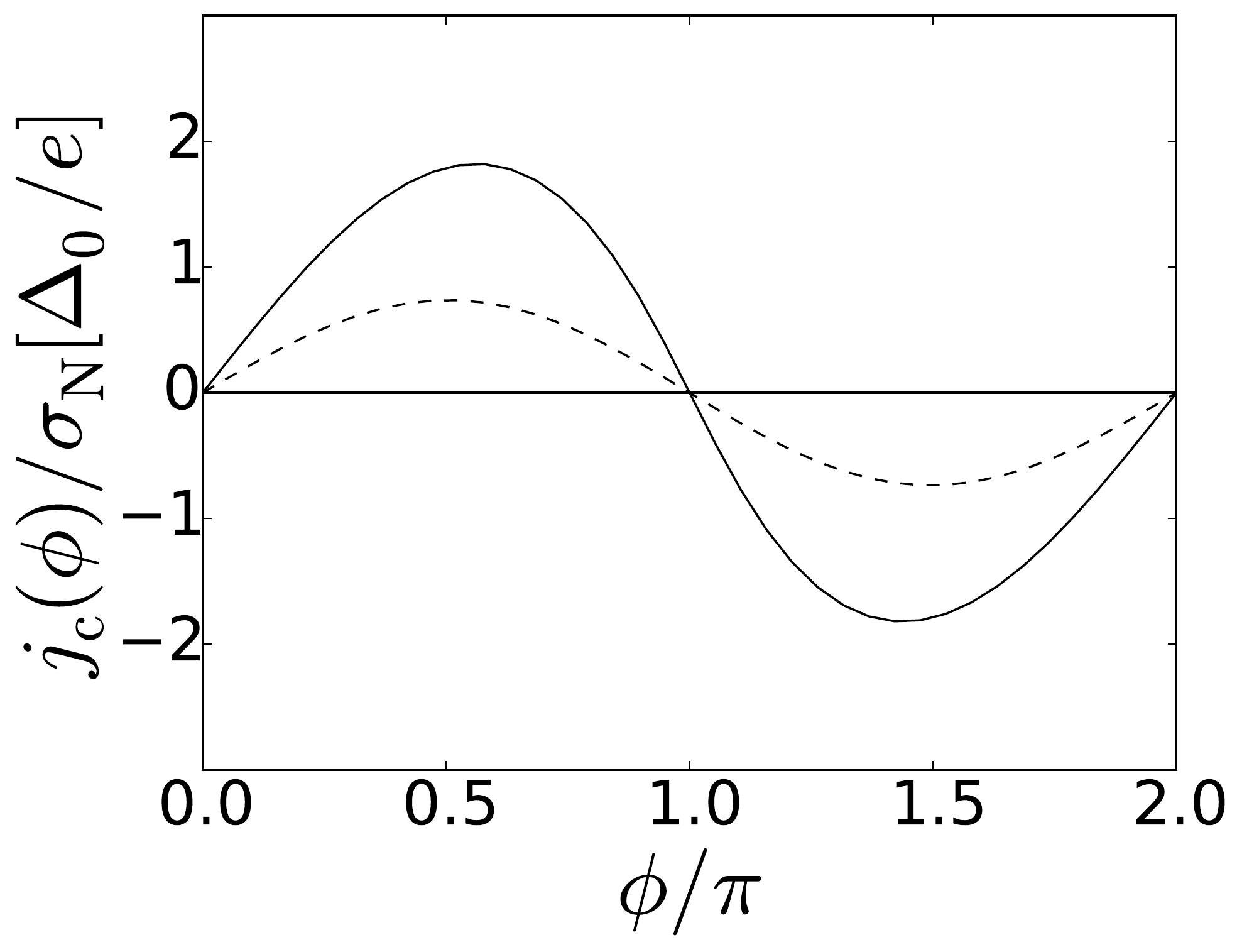}
	\includegraphics[width=.23\textwidth]{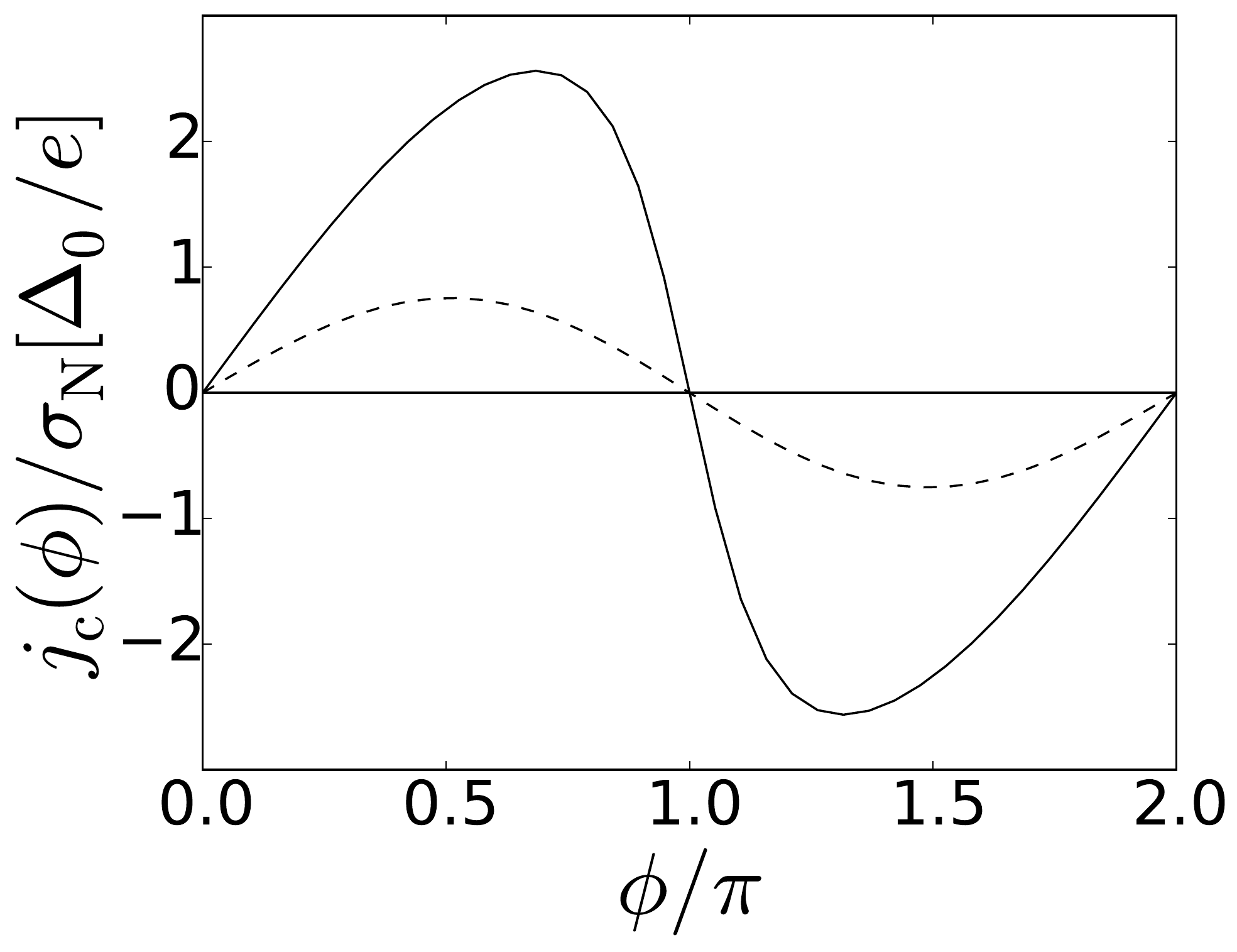}
	\includegraphics[width=.23\textwidth]{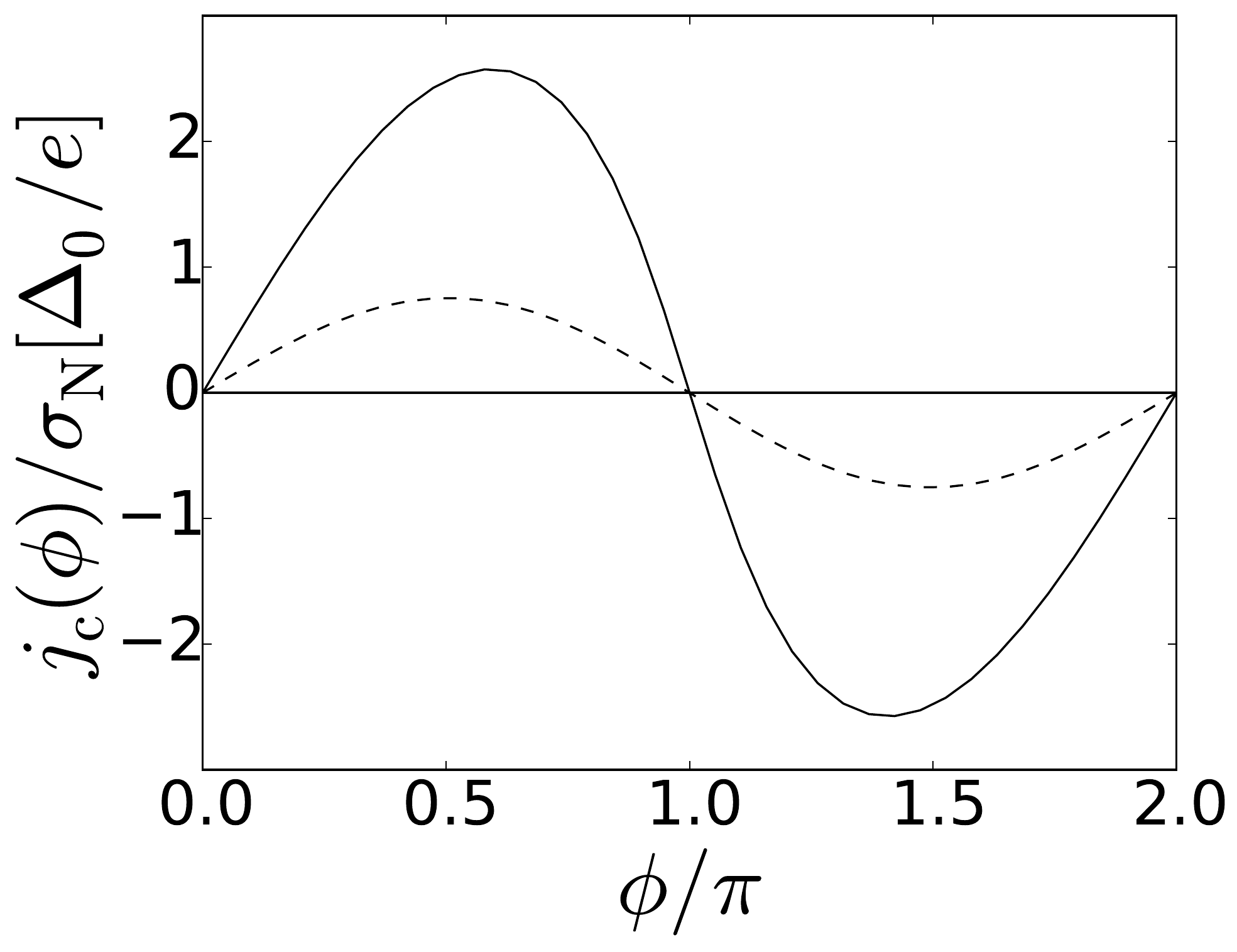}
	\caption{Current-phase relationship for the Josephson current density $j(\phi)$, normalized by the normal-state junction conductivity $\sigma_{\rm N}$. The top and bottom panels are for FFLO-like and BCS-like pairing, whereas the left and right panels are for the limits $|\mu| \gg \hbar v$ and $|\mu| \ll \hbar v$, respectively. The solid curves are for zero temperature; the dashed curves are for $k_{\rm B} T = \Delta_0$.}
	\label{fig:currentphase}
\end{figure} 

\begin{figure}[h!]
	\centering
	\includegraphics[width=.5\textwidth]{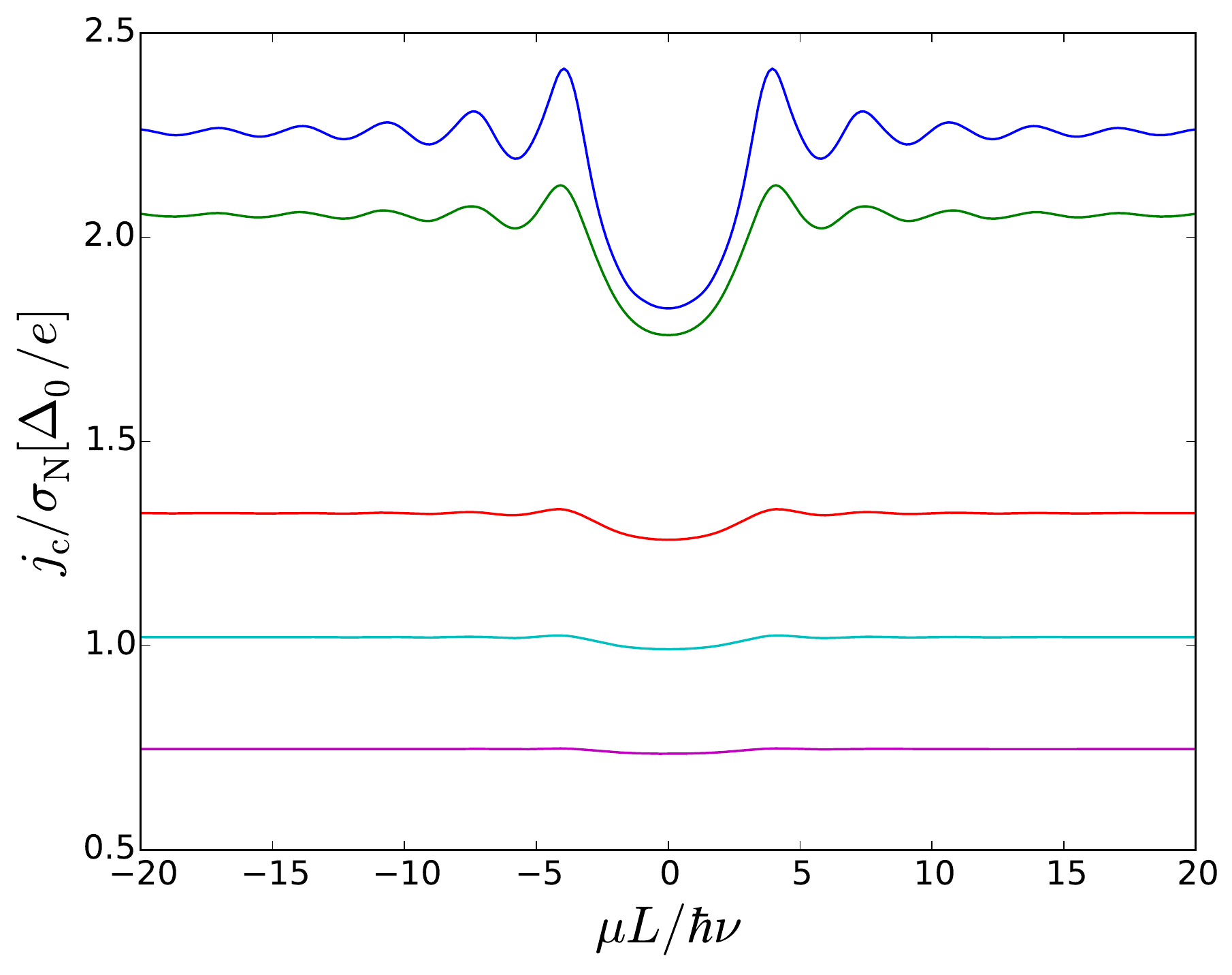}
	\includegraphics[width=.5\textwidth]{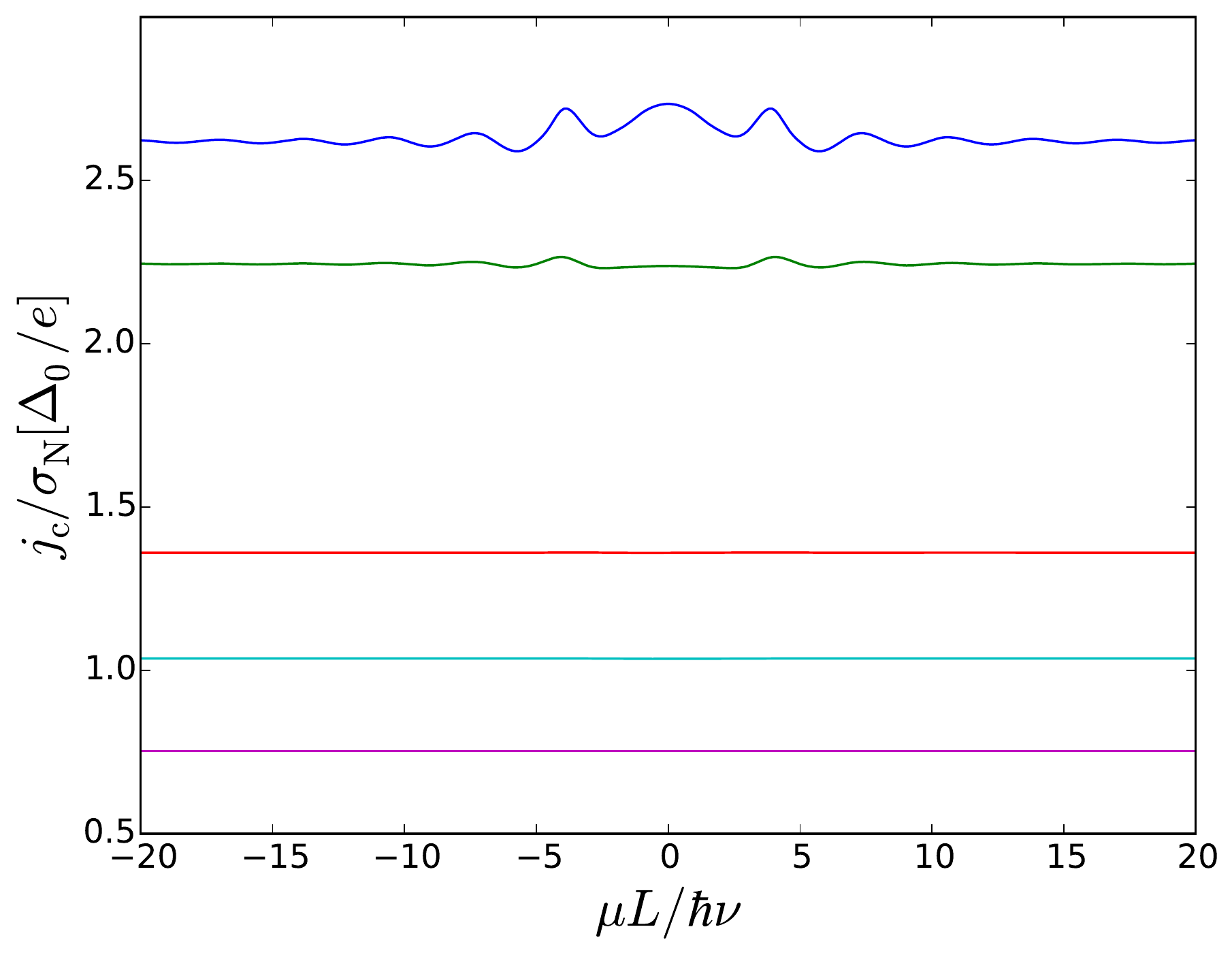}
	\caption{\label{fig:fig_ffloic}Critical current $j_{\rm c}$ normalized to normal-state junction conductivity $\sigma_{\rm N}$, as a function of $\mu L/\hbar v$. The top and bottom panels are for FFLO-like and BCS-like pairing, respectively. The different curves are for tempertures $k_{\rm B} T/\Delta_0 = 0$, $0.2$, $0.5$, $0.7$, and $1.0$ (top to bottom inside each panel).}
	\label{fig_ffloic}
\end{figure}

The Josephson current density $j(\phi)$ can be calculated from the results presented above and in the previous section. We were, however, not able to perform the integrations over the transverse momenta $q_x$ and $q_y$ in closed form, except for the high-temperature limit $k_{\rm B} T \gg \Delta_0$, for which we find
\begin{align}
  j_{\rm F,B}(\phi) = \frac{\Delta_{\rm F,B}^2 \sigma_{\rm N}}{2\pi e k_{\rm B} T} \sin{\phi} 
\end{align}
for FFLO-like pairing (F) and BCS-like pairing (B).
For temperatures $k_{\rm B} T \lesssim \Delta_0$, including zero temperature, we have to resort to a numerical evaluation of the Josephson current density. Figure \ref{fig:currentphase} shows the current-phase relationship for zero temperature and $k_{\rm B} T = \Delta_0$, normalized to the normal junction conductivity $\sigma_{\rm N}$, for the two limiting cases $|\mu|/\hbar v \ll 1$ and $|\mu|/\hbar v \gg 1$. Figure \ref{fig:fig_ffloic} shows the critical current density $j_{\rm c} = \max_{\phi} j(\phi)$, again normalized to the normal junction conductivity $\sigma_{\rm N}$, as a function of $\mu L/\hbar v$, {\em i.e.}, interpolating between the evanescent-mode dominated regime and the propagating-mode dominated regime.
Despite the vastly different conductivities in the two regimes [compare Eqs.\ (\ref{eq:sigmaN1}) and (\ref{eq:sigmaN2})], the ratio $j_{\rm c}/\sigma_{\rm N}$ shows a rather weak dependence on $\mu L/\hbar v$, especially in the case of BCS-like pairing and/or high temperatures. 

\section{Conclusion}\label{conclusions}

In this article, we have calculated the Josephson current through a thin normal-state Weyl (semi)metal slab. The superconducting terminals are modeled as strongly doped Weyl metals, and we considered both BCS-like and FFLO-like pairing mechanisms in the superconductors. The results for the FFLO-like pairing are qualitatively similar to results obtained previously by Titov and Beenakker for graphene, \cite{josephsongraphene1} up to numerical prefactors. The results for BCS-like pairing are quantitatively different. We find that the zero-temperature Josephson current density remains of order $\sigma_{\rm N} \Delta_0$ throughout the entire parameter range, where $\sigma_{\rm N}$ is the normal-state junction conductivity and $\Delta_0$ the superconducting gap.

A Weyl semimetal is known to have rather anomalous transport properties at the Weyl point $\mu=0$: A normal-state conductance that scales proportional to $W^2/L^2$ for a sample of width $W$ and length $L$, so that the conductance of a cubic sample is effectively size independent.\cite{Baireuther_PRB_89_035410,Sbierski_PRL_113_026602} (For comparison, in a conventional diffusive metal the conductance is proprortional to $L$ for a cubic sample in three dimensions.) Our at first sight somewhat unspectacular result that the Josephson current scales proportional to the normal-state conductivity throughout the entire parameter range we considered means that this anomalous scaling with system size $L$ also extends to the Josephson effect. This conclusion is consistent with the short-junction limit taken in our calculations, since the condition for evanescent-mode dominated transport is automatically satisfied in this limit for the typical energies of the Andreev bound states that mediate the Josephson current.

For the Weyl (semi)metal we have used a model with inversion symmetry, but with broken time-reversal symmetry, first proposed by Cho {\em et al.}\cite{weylsc4} Since then other models have been put forward, some of which have different symmetries for the (normal) Weyl semimetal or for the superconducting order, including possible magnetic phases.\cite{weylsc1,weylsc5,Meng_PRB_86_054504,Lu_PRL_114_096804} Another interesting outlook is to tilt the Weyl semimetal dispersion \cite{Bergholtz2015}—this is a generic feature of condensed matter Weyl realizations due to the lack of forbidding symmetries and is responsible for unusual transport characteristics.\cite{Trescher2015} A tilt of the spectrum also leads to an increased density of states which is favorable for the onset of superconductivity and, in fact, the previously mentioned compound MoTe$_2$ \cite{weylscexp} is a realization of an ''over-tilted'' so-called type-II Weyl semimetal.\cite{Soluyanov2015} It is a most intriguing question to extend the present calculations to the mentioned models possessing different symmetry properties, to see whether the Josephson effect can be used as an effective tool to distinguish between the different scenarios.

{\it Acknowledgements.} This work is supported by the Emmy Noether program (BE 5233/1-1) of the German Science Foundation DFG, the CRC TR 183 of the DFG (Project No. A02), the Swedish research council, and the Wallenberg Academy Fellows program of the Knut and Alice Wallenberg foundation.

\appendix

\section*{Appendix}

In this appendix we report the detailed calculation of the determinants $\det A_{\rm F}$ and $\det A_{\rm B}$ reported in Eqs.\ (\ref{eq:detAF}) and (\ref{eq:detAB}) of the main text.

Before calculating the scattering matrices, we first perform a gauge transformation that removes the large momentum $\vPW$ from the Bogoliubov-de Gennes equation. For the case of BCS-like pairing, this is the transformation 
$$
  H_{\rm B} \to \tilde H_{\rm B} = e^{i \vPW \cdot \vr} H_{\rm B} e^{-i \vPW \cdot \vr},
$$%
which results in the Bogoliubov-de Gennes Hamiltonian
\begin{equation}
	\tilde H_{\rm B} = 	\left(
	\begin{array}{cc}
		h_{+}(-i \vnabla)  - \mu & \Delta(z)^* \\
		\Delta(z) & \mu - h_{-}(-i \vnabla) \\
	\end{array}
	\right). \label{eq_bcsdbdg_app}
\end{equation}
For the FFLO-like pairing, we transform 
$$
  H_{\rm F,\pm} \to \tilde H_{\rm F,\pm} = e^{\pm i \sigma_z \vPW \cdot \vr} H_{\rm F} e^{\mp i \sigma_z \vPW \cdot \vr},
$$%
which gives
\begin{equation}
  \tilde H_{{\rm F},\pm} =
	\left(
	\begin{array}{cc}
		h_{\pm} (-i \vnabla) - \mu & \Delta(z) \\
		\Delta(z)^* & 
  \mu - h_{\pm} (-i \vnabla) \\
	\end{array}
	\right). \label{eq_fflodbdg_app}
\end{equation}

We also note that $h_{+}(-i \vnabla)$ is independent of the angle $\alpha$ between the momentum $\vPW$ of the Weyl point and the $z$ axis, whereas
\begin{equation}
  \sigma_x e^{i \alpha \sigma_y} h_{-}(-i \vnabla) e^{-i \alpha \sigma_y} \sigma_x = \overline{h}_{+}(-i \vnabla),
  \label{eq:rot}
\end{equation}
where $\overline{h}_{+}$ describes a Weyl node mirror-reflected in the $xz$ plane ({\em i.e.}, with $q_y \to -q_y$).
As a result, for FFLO-like pairing the two Weyl nodes will contribute equally to the supercurrent, and it is sufficient to calculate the contribution for $\tilde H_{{\rm F},+}$. For BCS-like pairing we perform the rotation (\ref{eq:rot}) for the hole block only, which leads to the transformed Bogoliubov-de Gennes Hamiltonian
\begin{equation}
  \tilde H_{{\rm B}} = \left(
	\begin{array}{cc}
		h_{+} (-i \vnabla) - \mu & \tilde \Delta(z) \\
		\tilde \Delta(z)^* &
  \mu - \overline{h}_{+} (-i \vnabla) \\
	\end{array}
	\right), \label{eq_fflodbdg_app_transf}
\end{equation}
with
\begin{equation}
  \tilde \Delta(z) = \Delta(z) \sigma_x \cos \alpha - \Delta(z) \sigma_z \sin \alpha.
\end{equation}
After this rotation the $\alpha$-dependence is moved to the superconducting order parameter $\tilde \Delta(z)$. In the limit $|V_0| \to \infty$ of a heavily doped superconductor the term proportional to $\sigma_x$ in the superconducting order parameter $\tilde \Delta(z)$ can be neglected, since it couples states with a large momentum difference. In this limit the replacement $\tilde \Delta(z) \rightarrow -\sigma_z \Delta_{\rm B}$ can be made.

To fix the basis for the calculation of the scattering matrices, we add an ``ideal lead'' --- a short segment of highly doped normal-state Weyl semimetal --- at $z = \pm L/2$. This segment is described by the normal-state Hamiltonians $h_{\pm}$, including the potential offset proportional to $V_0$. For this ideal lead, the flux-normalized spinors for left and right moving particle like states are
\begin{equation}
  u_{\rm R} = \left( \begin{array}{cc} 1 \\ 0 \end{array} \right),\ \
  u_{\rm L} = \left( \begin{array}{cc} 0 \\ 1 \end{array} \right).
\end{equation}
We have suppressed the dependence on the transverse coordinates $x$ and $y$, which is $\propto e^{i q_x x + i q_y y}$ throughout. Similarly, the spinors for hole like states are
\begin{align}
  v_{\rm R} =& \left( \begin{array}{cc} 0 \\ 1 \end{array} \right), \ \
  v_{\rm L} = \left( \begin{array}{cc} 1 \\ 0 \end{array} \right). 
\end{align}

Since we are interested in the short-junction limit only, it is sufficient to find the scattering matrix of the normal region $-L/2 < z < L/z$ for energy $\varepsilon = 0$. Hereto we parametrize
\begin{equation}
  q = \frac{\mu}{\hbar v},\ \ q_{\perp} = \sqrt{q_x^2 + q_y^2},
\end{equation}
and
\begin{equation}
  q_x = q \sin \theta \cos \varphi,\ \
  q_y = q \sin \theta \sin \varphi,\ \
  q_z = q \cos \theta.
\end{equation}
Note that $\theta$ is complex if $q_x^2 + q_y^2 > q^2$.
The electron wave function at the Weyl point $\vPW$ is a linear combination of the two (unnormalized) basis states,
\begin{align*}
  & \left( \begin{array}{c} \sin(\theta/2) \\
  e^{i \varphi} \cos(\theta/2) \end{array} \right) e^{-i q_z z} \ 
  \mbox{,} \ 
  \left( \begin{array}{c} \cos(\theta/2) \\
  e^{i \varphi} \sin(\theta/2) \end{array} \right) e^{i q_z z}.
\end{align*}
Imposing continuity of the wave function at $z = -L/2$ and $z = L/2$ we then find the transmission and reflection amplitudes,
\begin{align}
  s_{\pm \pm}(\varphi,\theta) &=
  \frac{e^{\mp i \varphi} q_{\perp} \sin(q_z l)}{q \sin(q_z l) + i q_z \cos(q_z L)}, \nonumber \\
  s_{\pm \mp}(\varphi,\theta) &= 
  \frac{i q_z}{q \sin(q_z L) + i q_z \cos(q_z L)}.
\end{align}
The normal-state transmission probability $T(q_x,q_y)$ of Eq.\ (\ref{eq:TN}) is $T(q_x,q_y) = |s_{+-}(q_x,q_y)|^2$.
For FFLO-like pairing the reflection and transmission amplitudes for holes are the complex conjugates of the reflection and transmission amplitudes for the electrons, so that we find
\begin{align}
  {\cal S}_{\tau\tau'}(\varphi,\theta) = \left( \begin{array}{cc} s_{\tau\tau'}(\varphi,\theta) & 0 \\
  0 & s_{\tau \tau'}(\varphi,\theta)^* \end{array} \right),
  \label{eq:SNormalF}
\end{align}
for $\tau,\tau'=\pm$. For BCS-like pairing an additional reflection in the $xz$ plane has to be performed for the reflection and transmission amplitudes for the holes,
\begin{align}
  {\cal S}_{\tau\tau'}(\varphi,\theta) = \left( \begin{array}{cc} s_{\tau\tau'}(\varphi,\theta) & 0 \\
  0 & s_{\tau \tau'}(-\varphi,\theta)^* \end{array} \right),
  \label{eq:SNormalB}
\end{align}

We now calculate the reflection matrices for reflection off the heavily doped superconducting regions $z < -L/2$ and $z > L/2$. We first consider the case of FFLO-like pairing. Abbreviating 
\begin{equation}
  q_0 = \frac{\mu - V_0}{\hbar v},
\end{equation}
and taking the limit $V_0 \to -\infty$, the basis spinors for $z > L/2$ take the simple form,
$$
  \left( \begin{array}{c} e^{i \phi/2} \\ 0 \\ e^{i \gamma_{\rm F}} \\ 0 \end{array} \right) e^{i q_0 z - \kappa_{\rm F} z}
 \ \mbox{,} \
  \left( \begin{array}{c} 0 \\ e^{i \phi/2} \\ 0 \\ e^{-i \gamma_{\rm F}} \end{array} \right) e^{-i q_0 z - \kappa_{\rm F} z},
$$%
where 
\begin{equation}
  \gamma_{\rm F} = -\arccos \frac{\varepsilon}{\Delta_{\rm F}},\ \
  \kappa_{\rm F} = \frac{1}{\hbar v} \sqrt{\Delta_{\rm F}^2 - \varepsilon^2}.
\end{equation}
For $z < -L/2$ the basis spinors are
$$
  \left( \begin{array}{c} e^{-i \phi/2} \\ 0 \\ e^{-i \gamma_{\rm F}} \\ 0 \end{array} \right) e^{i q_0 z + \kappa_{\rm F} z}
 \ \mbox{,} \
  \left( \begin{array}{c} 0 \\ e^{-i \phi/2} \\ 0 \\ e^{i \gamma_{\rm F}} \end{array} \right) e^{-i q_0 z + \kappa_{\rm F} z}.
$$%
For BCS-like pairing the last (fourth) entry of these basis vectors has to be multiplied by $-1$ and the coefficients $\gamma_{\rm F}$ and $\kappa_{\rm F}$ have to be replaced by $\gamma_{\rm B}$ and $\kappa_{\rm B}$, respectively, with
\begin{equation}
  \gamma_{\rm B} = -\arccos\frac{\varepsilon}{\Delta_{\rm B}},\ \
  \kappa_{\rm B} = \frac{1}{\hbar v} \sqrt{\Delta_{\rm B}^2 - \varepsilon^2}.
\end{equation}
Matching wave functions at the interfaces at $z=\pm L/2$ we find the reflection matrices,
\begin{align}
  {\cal R}_+ &= \left( \begin{array}{cc} 0 & e^{i \gamma + i \phi/2} \\ e^{i \gamma - i \phi/2} & 0 \end{array} \right), \nonumber \\
  {\cal R}_- &= \left( \begin{array}{cc} 0 & e^{i \gamma - i \phi/2} \\ e^{i \gamma + i \phi/2} & 0 \end{array} \right),
\end{align}
for FFLO-like pairing and
\begin{align}
  {\cal R}_+ &= \left( \begin{array}{cc} 0 & -e^{i \gamma + i \phi/2} \\ e^{i \gamma - i \phi/2} & 0 \end{array} \right), \nonumber \\
  {\cal R}_- &= \left( \begin{array}{cc} 0 & e^{i \gamma - i \phi/2} \\ - e^{i \gamma + i \phi/2} & 0 \end{array} \right),
\end{align}
for BCS-like pairing. Note that these matrices represent perfect Andreev reflection. Normal reflection at the superconducting interfaces is included in the normal-layer scattering matrices (\ref{eq:SNormalF}) and (\ref{eq:SNormalB}).

We then find that
\begin{align}
  \det A_{\rm F} =&\, 
  1 + e^{4 i \gamma_{\rm F}} -
    2 e^{2 i \gamma_{\rm F}}
  \frac{q_z^2 \cos \phi + q_{\perp}^2 \sin^2 (q_z L)}{q_z^2 + q_{\perp}^2 \sin^2 (q_z L)}, \\
  \det A_{\rm B} =&\, 
  1 + e^{4 i \gamma_{\rm B}} \nonumber \\ &\, \mbox{}
   - 2 e^{2 i \gamma_{\rm B}}
  \frac{q_z^2 \cos \phi - q_{\perp}^2 \cos(2 \varphi) \sin^2(q_z L)}{q_z^2 + q_{\perp}^2 \sin^2 (q_z L)},
\end{align}
which is the same as the results (\ref{eq:detAF}) and (\ref{eq:detAB}) of the main text.

\bibliography{bibliography}

\begin{thebibliography}{54}%
\makeatletter
\providecommand \@ifxundefined [1]{%
 \@ifx{#1\undefined}
}%
\providecommand \@ifnum [1]{%
 \ifnum #1\expandafter \@firstoftwo
 \else \expandafter \@secondoftwo
 \fi
}%
\providecommand \@ifx [1]{%
 \ifx #1\expandafter \@firstoftwo
 \else \expandafter \@secondoftwo
 \fi
}%
\providecommand \natexlab [1]{#1}%
\providecommand \enquote  [1]{``#1''}%
\providecommand \bibnamefont  [1]{#1}%
\providecommand \bibfnamefont [1]{#1}%
\providecommand \citenamefont [1]{#1}%
\providecommand \href@noop [0]{\@secondoftwo}%
\providecommand \href [0]{\begingroup \@sanitize@url \@href}%
\providecommand \@href[1]{\@@startlink{#1}\@@href}%
\providecommand \@@href[1]{\endgroup#1\@@endlink}%
\providecommand \@sanitize@url [0]{\catcode `\\12\catcode `\$12\catcode
  `\&12\catcode `\#12\catcode `\^12\catcode `\_12\catcode `\%12\relax}%
\providecommand \@@startlink[1]{}%
\providecommand \@@endlink[0]{}%
\providecommand \url  [0]{\begingroup\@sanitize@url \@url }%
\providecommand \@url [1]{\endgroup\@href {#1}{\urlprefix }}%
\providecommand \urlprefix  [0]{URL }%
\providecommand \Eprint [0]{\href }%
\providecommand \doibase [0]{http://dx.doi.org/}%
\providecommand \selectlanguage [0]{\@gobble}%
\providecommand \bibinfo  [0]{\@secondoftwo}%
\providecommand \bibfield  [0]{\@secondoftwo}%
\providecommand \translation [1]{[#1]}%
\providecommand \BibitemOpen [0]{}%
\providecommand \bibitemStop [0]{}%
\providecommand \bibitemNoStop [0]{.\EOS\space}%
\providecommand \EOS [0]{\spacefactor3000\relax}%
\providecommand \BibitemShut  [1]{\csname bibitem#1\endcsname}%
\let\auto@bib@innerbib\@empty
\bibitem [{\citenamefont {{Turner}}\ and\ \citenamefont
  {{Vishwanath}}()}]{weylturnerreview}%
  \BibitemOpen
  \bibfield  {author} {\bibinfo {author} {\bibfnamefont {A.~M.}\ \bibnamefont
  {{Turner}}}\ and\ \bibinfo {author} {\bibfnamefont {A.}~\bibnamefont
  {{Vishwanath}}},\ }\href@noop {} {\ }\Eprint {http://arxiv.org/abs/1301.0330}
  {arXiv:1301.0330} \BibitemShut {NoStop}%
\bibitem [{\citenamefont {Hosur}\ and\ \citenamefont
  {Qi}(2013)}]{weylhosurreview}%
  \BibitemOpen
  \bibfield  {author} {\bibinfo {author} {\bibfnamefont {P.}~\bibnamefont
  {Hosur}}\ and\ \bibinfo {author} {\bibfnamefont {X.}~\bibnamefont {Qi}},\
  }\href {\doibase http://dx.doi.org/10.1016/j.crhy.2013.10.010} {\bibfield
  {journal} {\bibinfo  {journal} {C. R. Phys.}\ }\textbf {\bibinfo {volume}
  {14}},\ \bibinfo {pages} {857 } (\bibinfo {year} {2013})}\BibitemShut
  {NoStop}%
\bibitem [{\citenamefont {Xu}\ \emph {et~al.}(2015)\citenamefont {Xu},
  \citenamefont {Liu}, \citenamefont {Kushwaha}, \citenamefont {Sankar},
  \citenamefont {Krizan}, \citenamefont {Belopolski}, \citenamefont {Neupane},
  \citenamefont {Bian}, \citenamefont {Alidoust}, \citenamefont {Chang},
  \citenamefont {Jeng}, \citenamefont {Huang}, \citenamefont {Tsai},
  \citenamefont {Lin}, \citenamefont {Shibayev}, \citenamefont {Chou},
  \citenamefont {Cava},\ and\ \citenamefont {Hasan}}]{weylexp1}%
  \BibitemOpen
  \bibfield  {author} {\bibinfo {author} {\bibfnamefont {S.-Y.}\ \bibnamefont
  {Xu}}, \bibinfo {author} {\bibfnamefont {C.}~\bibnamefont {Liu}}, \bibinfo
  {author} {\bibfnamefont {S.~K.}\ \bibnamefont {Kushwaha}}, \bibinfo {author}
  {\bibfnamefont {R.}~\bibnamefont {Sankar}}, \bibinfo {author} {\bibfnamefont
  {J.~W.}\ \bibnamefont {Krizan}}, \bibinfo {author} {\bibfnamefont
  {I.}~\bibnamefont {Belopolski}}, \bibinfo {author} {\bibfnamefont
  {M.}~\bibnamefont {Neupane}}, \bibinfo {author} {\bibfnamefont
  {G.}~\bibnamefont {Bian}}, \bibinfo {author} {\bibfnamefont {N.}~\bibnamefont
  {Alidoust}}, \bibinfo {author} {\bibfnamefont {T.-R.}\ \bibnamefont {Chang}},
  \bibinfo {author} {\bibfnamefont {H.-T.}\ \bibnamefont {Jeng}}, \bibinfo
  {author} {\bibfnamefont {C.-Y.}\ \bibnamefont {Huang}}, \bibinfo {author}
  {\bibfnamefont {W.-F.}\ \bibnamefont {Tsai}}, \bibinfo {author}
  {\bibfnamefont {H.}~\bibnamefont {Lin}}, \bibinfo {author} {\bibfnamefont
  {P.~P.}\ \bibnamefont {Shibayev}}, \bibinfo {author} {\bibfnamefont {F.-C.}\
  \bibnamefont {Chou}}, \bibinfo {author} {\bibfnamefont {R.~J.}\ \bibnamefont
  {Cava}}, \ and\ \bibinfo {author} {\bibfnamefont {M.~Z.}\ \bibnamefont
  {Hasan}},\ }\href {\doibase 10.1126/science.1256742} {\bibfield  {journal}
  {\bibinfo  {journal} {Science}\ }\textbf {\bibinfo {volume} {347}},\ \bibinfo
  {pages} {294} (\bibinfo {year} {2015})}\BibitemShut {NoStop}%
\bibitem [{\citenamefont {{Zhang}}\ \emph {et~al.}()\citenamefont {{Zhang}},
  \citenamefont {{Lin}}, \citenamefont {{Guo}}, \citenamefont {{Xu}},
  \citenamefont {{Lee}}, \citenamefont {{Lu}}, \citenamefont {{Huang}},
  \citenamefont {{Chang}}, \citenamefont {{Hsu}}, \citenamefont {{Lin}},
  \citenamefont {{Li}}, \citenamefont {{Zhang}}, \citenamefont {{Neupert}},
  \citenamefont {{Z. Hasan}}, \citenamefont {{Wang}},\ and\ \citenamefont
  {{Jia}}}]{weylexp2}%
  \BibitemOpen
  \bibfield  {author} {\bibinfo {author} {\bibfnamefont {C.}~\bibnamefont
  {{Zhang}}}, \bibinfo {author} {\bibfnamefont {Z.}~\bibnamefont {{Lin}}},
  \bibinfo {author} {\bibfnamefont {C.}~\bibnamefont {{Guo}}}, \bibinfo
  {author} {\bibfnamefont {S.-Y.}\ \bibnamefont {{Xu}}}, \bibinfo {author}
  {\bibfnamefont {C.-C.}\ \bibnamefont {{Lee}}}, \bibinfo {author}
  {\bibfnamefont {H.}~\bibnamefont {{Lu}}}, \bibinfo {author} {\bibfnamefont
  {S.-M.}\ \bibnamefont {{Huang}}}, \bibinfo {author} {\bibfnamefont
  {G.}~\bibnamefont {{Chang}}}, \bibinfo {author} {\bibfnamefont {C.-H.}\
  \bibnamefont {{Hsu}}}, \bibinfo {author} {\bibfnamefont {H.}~\bibnamefont
  {{Lin}}}, \bibinfo {author} {\bibfnamefont {L.}~\bibnamefont {{Li}}},
  \bibinfo {author} {\bibfnamefont {C.}~\bibnamefont {{Zhang}}}, \bibinfo
  {author} {\bibfnamefont {T.}~\bibnamefont {{Neupert}}}, \bibinfo {author}
  {\bibfnamefont {M.}~\bibnamefont {{Z. Hasan}}}, \bibinfo {author}
  {\bibfnamefont {J.}~\bibnamefont {{Wang}}}, \ and\ \bibinfo {author}
  {\bibfnamefont {S.}~\bibnamefont {{Jia}}},\ }\href@noop {} {\ }\Eprint
  {http://arxiv.org/abs/1507.06301} {arXiv:1507.06301} \BibitemShut {NoStop}%
\bibitem [{\citenamefont {{Xu}}\ \emph {et~al.}(2015)\citenamefont {{Xu}},
  \citenamefont {{Belopolski}}, \citenamefont {{Sanchez}}, \citenamefont
  {{Zhang}}, \citenamefont {{Chang}}, \citenamefont {{Guo}}, \citenamefont
  {{Bian}}, \citenamefont {{Yuan}}, \citenamefont {{Lu}}, \citenamefont
  {{Chang}}, \citenamefont {{Shibayev}}, \citenamefont {{Prokopovych}},
  \citenamefont {{Alidoust}}, \citenamefont {{Zheng}}, \citenamefont {{Lee}},
  \citenamefont {{Huang}}, \citenamefont {{Sankar}}, \citenamefont {{Chou}},
  \citenamefont {{Hsu}}, \citenamefont {{Jeng}}, \citenamefont {{Bansil}},
  \citenamefont {{Neupert}}, \citenamefont {{Strocov}}, \citenamefont {{Lin}},
  \citenamefont {{Jia}},\ and\ \citenamefont {{Hasan}}}]{weylexp3}%
  \BibitemOpen
  \bibfield  {author} {\bibinfo {author} {\bibfnamefont {S.-Y.}\ \bibnamefont
  {{Xu}}}, \bibinfo {author} {\bibfnamefont {I.}~\bibnamefont {{Belopolski}}},
  \bibinfo {author} {\bibfnamefont {D.~S.}\ \bibnamefont {{Sanchez}}}, \bibinfo
  {author} {\bibfnamefont {C.}~\bibnamefont {{Zhang}}}, \bibinfo {author}
  {\bibfnamefont {G.}~\bibnamefont {{Chang}}}, \bibinfo {author} {\bibfnamefont
  {C.}~\bibnamefont {{Guo}}}, \bibinfo {author} {\bibfnamefont
  {G.}~\bibnamefont {{Bian}}}, \bibinfo {author} {\bibfnamefont
  {Z.}~\bibnamefont {{Yuan}}}, \bibinfo {author} {\bibfnamefont
  {H.}~\bibnamefont {{Lu}}}, \bibinfo {author} {\bibfnamefont {T.-R.}\
  \bibnamefont {{Chang}}}, \bibinfo {author} {\bibfnamefont {P.~P.}\
  \bibnamefont {{Shibayev}}}, \bibinfo {author} {\bibfnamefont {M.~L.}\
  \bibnamefont {{Prokopovych}}}, \bibinfo {author} {\bibfnamefont
  {N.}~\bibnamefont {{Alidoust}}}, \bibinfo {author} {\bibfnamefont
  {H.}~\bibnamefont {{Zheng}}}, \bibinfo {author} {\bibfnamefont {C.-C.}\
  \bibnamefont {{Lee}}}, \bibinfo {author} {\bibfnamefont {S.-M.}\ \bibnamefont
  {{Huang}}}, \bibinfo {author} {\bibfnamefont {R.}~\bibnamefont {{Sankar}}},
  \bibinfo {author} {\bibfnamefont {F.}~\bibnamefont {{Chou}}}, \bibinfo
  {author} {\bibfnamefont {C.-H.}\ \bibnamefont {{Hsu}}}, \bibinfo {author}
  {\bibfnamefont {H.-T.}\ \bibnamefont {{Jeng}}}, \bibinfo {author}
  {\bibfnamefont {A.}~\bibnamefont {{Bansil}}}, \bibinfo {author}
  {\bibfnamefont {T.}~\bibnamefont {{Neupert}}}, \bibinfo {author}
  {\bibfnamefont {V.~N.}\ \bibnamefont {{Strocov}}}, \bibinfo {author}
  {\bibfnamefont {H.}~\bibnamefont {{Lin}}}, \bibinfo {author} {\bibfnamefont
  {S.}~\bibnamefont {{Jia}}}, \ and\ \bibinfo {author} {\bibfnamefont {M.~Z.}\
  \bibnamefont {{Hasan}}},\ }\href {\doibase 10.1126/sciadv.1501092} {\bibfield
   {journal} {\bibinfo  {journal} {Sci. Adv.}\ }\textbf {\bibinfo {volume}
  {1}},\ \bibinfo {pages} {e1501092} (\bibinfo {year} {2015})}\BibitemShut
  {NoStop}%
\bibitem [{\citenamefont {Xu}\ \emph {et~al.}(2015)\citenamefont {Xu},
  \citenamefont {Belopolski}, \citenamefont {Alidoust}, \citenamefont
  {Neupane}, \citenamefont {Bian}, \citenamefont {Zhang}, \citenamefont
  {Sankar}, \citenamefont {Chang}, \citenamefont {Yuan}, \citenamefont {Lee},
  \citenamefont {Huang}, \citenamefont {Zheng}, \citenamefont {Ma},
  \citenamefont {Sanchez}, \citenamefont {Wang}, \citenamefont {Bansil},
  \citenamefont {Chou}, \citenamefont {Shibayev}, \citenamefont {Lin},
  \citenamefont {Jia},\ and\ \citenamefont {Hasan}}]{weylexp4}%
  \BibitemOpen
  \bibfield  {author} {\bibinfo {author} {\bibfnamefont {S.-Y.}\ \bibnamefont
  {Xu}}, \bibinfo {author} {\bibfnamefont {I.}~\bibnamefont {Belopolski}},
  \bibinfo {author} {\bibfnamefont {N.}~\bibnamefont {Alidoust}}, \bibinfo
  {author} {\bibfnamefont {M.}~\bibnamefont {Neupane}}, \bibinfo {author}
  {\bibfnamefont {G.}~\bibnamefont {Bian}}, \bibinfo {author} {\bibfnamefont
  {C.}~\bibnamefont {Zhang}}, \bibinfo {author} {\bibfnamefont
  {R.}~\bibnamefont {Sankar}}, \bibinfo {author} {\bibfnamefont
  {G.}~\bibnamefont {Chang}}, \bibinfo {author} {\bibfnamefont
  {Z.}~\bibnamefont {Yuan}}, \bibinfo {author} {\bibfnamefont {C.-C.}\
  \bibnamefont {Lee}}, \bibinfo {author} {\bibfnamefont {S.-M.}\ \bibnamefont
  {Huang}}, \bibinfo {author} {\bibfnamefont {H.}~\bibnamefont {Zheng}},
  \bibinfo {author} {\bibfnamefont {J.}~\bibnamefont {Ma}}, \bibinfo {author}
  {\bibfnamefont {D.~S.}\ \bibnamefont {Sanchez}}, \bibinfo {author}
  {\bibfnamefont {B.}~\bibnamefont {Wang}}, \bibinfo {author} {\bibfnamefont
  {A.}~\bibnamefont {Bansil}}, \bibinfo {author} {\bibfnamefont
  {F.}~\bibnamefont {Chou}}, \bibinfo {author} {\bibfnamefont {P.~P.}\
  \bibnamefont {Shibayev}}, \bibinfo {author} {\bibfnamefont {H.}~\bibnamefont
  {Lin}}, \bibinfo {author} {\bibfnamefont {S.}~\bibnamefont {Jia}}, \ and\
  \bibinfo {author} {\bibfnamefont {M.~Z.}\ \bibnamefont {Hasan}},\ }\href
  {\doibase 10.1126/science.aaa9297} {\bibfield  {journal} {\bibinfo  {journal}
  {Science}\ }\textbf {\bibinfo {volume} {349}},\ \bibinfo {pages} {613}
  (\bibinfo {year} {2015})}\BibitemShut {NoStop}%
\bibitem [{\citenamefont {Lv}\ \emph {et~al.}(2015)\citenamefont {Lv},
  \citenamefont {Weng}, \citenamefont {Fu}, \citenamefont {Wang}, \citenamefont
  {Miao}, \citenamefont {Ma}, \citenamefont {Richard}, \citenamefont {Huang},
  \citenamefont {Zhao}, \citenamefont {Chen}, \citenamefont {Fang},
  \citenamefont {Dai}, \citenamefont {Qian},\ and\ \citenamefont
  {Ding}}]{weylexp5}%
  \BibitemOpen
  \bibfield  {author} {\bibinfo {author} {\bibfnamefont {B.~Q.}\ \bibnamefont
  {Lv}}, \bibinfo {author} {\bibfnamefont {H.~M.}\ \bibnamefont {Weng}},
  \bibinfo {author} {\bibfnamefont {B.~B.}\ \bibnamefont {Fu}}, \bibinfo
  {author} {\bibfnamefont {X.~P.}\ \bibnamefont {Wang}}, \bibinfo {author}
  {\bibfnamefont {H.}~\bibnamefont {Miao}}, \bibinfo {author} {\bibfnamefont
  {J.}~\bibnamefont {Ma}}, \bibinfo {author} {\bibfnamefont {P.}~\bibnamefont
  {Richard}}, \bibinfo {author} {\bibfnamefont {X.~C.}\ \bibnamefont {Huang}},
  \bibinfo {author} {\bibfnamefont {L.~X.}\ \bibnamefont {Zhao}}, \bibinfo
  {author} {\bibfnamefont {G.~F.}\ \bibnamefont {Chen}}, \bibinfo {author}
  {\bibfnamefont {Z.}~\bibnamefont {Fang}}, \bibinfo {author} {\bibfnamefont
  {X.}~\bibnamefont {Dai}}, \bibinfo {author} {\bibfnamefont {T.}~\bibnamefont
  {Qian}}, \ and\ \bibinfo {author} {\bibfnamefont {H.}~\bibnamefont {Ding}},\
  }\href {\doibase 10.1103/PhysRevX.5.031013} {\bibfield  {journal} {\bibinfo
  {journal} {Phys. Rev. X}\ }\textbf {\bibinfo {volume} {5}},\ \bibinfo {pages}
  {031013} (\bibinfo {year} {2015})}\BibitemShut {NoStop}%
\bibitem [{\citenamefont {{Lv}}\ \emph {et~al.}(2015)\citenamefont {{Lv}},
  \citenamefont {{Xu}}, \citenamefont {{Weng}}, \citenamefont {{Ma}},
  \citenamefont {{Richard}}, \citenamefont {{Huang}}, \citenamefont {{Zhao}},
  \citenamefont {{Chen}}, \citenamefont {{Matt}}, \citenamefont {{Bisti}},
  \citenamefont {{Strocov}}, \citenamefont {{Mesot}}, \citenamefont {{Fang}},
  \citenamefont {{Dai}}, \citenamefont {{Qian}}, \citenamefont {{Shi}},\ and\
  \citenamefont {{Ding}}}]{weylexp6}%
  \BibitemOpen
  \bibfield  {author} {\bibinfo {author} {\bibfnamefont {B.~Q.}\ \bibnamefont
  {{Lv}}}, \bibinfo {author} {\bibfnamefont {N.}~\bibnamefont {{Xu}}}, \bibinfo
  {author} {\bibfnamefont {H.~M.}\ \bibnamefont {{Weng}}}, \bibinfo {author}
  {\bibfnamefont {J.~Z.}\ \bibnamefont {{Ma}}}, \bibinfo {author}
  {\bibfnamefont {P.}~\bibnamefont {{Richard}}}, \bibinfo {author}
  {\bibfnamefont {X.~C.}\ \bibnamefont {{Huang}}}, \bibinfo {author}
  {\bibfnamefont {L.~X.}\ \bibnamefont {{Zhao}}}, \bibinfo {author}
  {\bibfnamefont {G.~F.}\ \bibnamefont {{Chen}}}, \bibinfo {author}
  {\bibfnamefont {C.~E.}\ \bibnamefont {{Matt}}}, \bibinfo {author}
  {\bibfnamefont {F.}~\bibnamefont {{Bisti}}}, \bibinfo {author} {\bibfnamefont
  {V.~N.}\ \bibnamefont {{Strocov}}}, \bibinfo {author} {\bibfnamefont
  {J.}~\bibnamefont {{Mesot}}}, \bibinfo {author} {\bibfnamefont
  {Z.}~\bibnamefont {{Fang}}}, \bibinfo {author} {\bibfnamefont
  {X.}~\bibnamefont {{Dai}}}, \bibinfo {author} {\bibfnamefont
  {T.}~\bibnamefont {{Qian}}}, \bibinfo {author} {\bibfnamefont
  {M.}~\bibnamefont {{Shi}}}, \ and\ \bibinfo {author} {\bibfnamefont
  {H.}~\bibnamefont {{Ding}}},\ }\href {\doibase 10.1038/nphys3426} {\bibfield
  {journal} {\bibinfo  {journal} {Nat. Phys.}\ }\textbf {\bibinfo {volume}
  {11}},\ \bibinfo {pages} {724} (\bibinfo {year} {2015})}\BibitemShut
  {NoStop}%
\bibitem [{\citenamefont {{Yang}}\ \emph {et~al.}(2015)\citenamefont {{Yang}},
  \citenamefont {{Liu}}, \citenamefont {{Sun}}, \citenamefont {{Peng}},
  \citenamefont {{Yang}}, \citenamefont {{Zhang}}, \citenamefont {{Zhou}},
  \citenamefont {{Zhang}}, \citenamefont {{Guo}}, \citenamefont {{Rahn}},
  \citenamefont {{Prabhakaran}}, \citenamefont {{Hussain}}, \citenamefont
  {{Mo}}, \citenamefont {{Felser}}, \citenamefont {{Yan}},\ and\ \citenamefont
  {{Chen}}}]{weylexp7}%
  \BibitemOpen
  \bibfield  {author} {\bibinfo {author} {\bibfnamefont {L.}~\bibnamefont
  {{Yang}}}, \bibinfo {author} {\bibfnamefont {Z.}~\bibnamefont {{Liu}}},
  \bibinfo {author} {\bibfnamefont {Y.}~\bibnamefont {{Sun}}}, \bibinfo
  {author} {\bibfnamefont {H.}~\bibnamefont {{Peng}}}, \bibinfo {author}
  {\bibfnamefont {H.}~\bibnamefont {{Yang}}}, \bibinfo {author} {\bibfnamefont
  {T.}~\bibnamefont {{Zhang}}}, \bibinfo {author} {\bibfnamefont
  {B.}~\bibnamefont {{Zhou}}}, \bibinfo {author} {\bibfnamefont
  {Y.}~\bibnamefont {{Zhang}}}, \bibinfo {author} {\bibfnamefont
  {Y.}~\bibnamefont {{Guo}}}, \bibinfo {author} {\bibfnamefont
  {M.}~\bibnamefont {{Rahn}}}, \bibinfo {author} {\bibfnamefont
  {D.}~\bibnamefont {{Prabhakaran}}}, \bibinfo {author} {\bibfnamefont
  {Z.}~\bibnamefont {{Hussain}}}, \bibinfo {author} {\bibfnamefont {S.-K.}\
  \bibnamefont {{Mo}}}, \bibinfo {author} {\bibfnamefont {C.}~\bibnamefont
  {{Felser}}}, \bibinfo {author} {\bibfnamefont {B.}~\bibnamefont {{Yan}}}, \
  and\ \bibinfo {author} {\bibfnamefont {Y.}~\bibnamefont {{Chen}}},\ }\href
  {http://dx.doi.org/10.1038/nphys3425} {\bibfield  {journal} {\bibinfo
  {journal} {Nat. Phys.}\ }\textbf {\bibinfo {volume} {11}},\ \bibinfo {pages}
  {728} (\bibinfo {year} {2015})}\BibitemShut {NoStop}%
\bibitem [{\citenamefont {{Borisenko}}\ \emph {et~al.}()\citenamefont
  {{Borisenko}}, \citenamefont {{Evtushinsky}}, \citenamefont {{Gibson}},
  \citenamefont {{Yaresko}}, \citenamefont {{Kim}}, \citenamefont {{Ali}},
  \citenamefont {{Buechner}}, \citenamefont {{Hoesch}},\ and\ \citenamefont
  {{Cava}}}]{weylexp8}%
  \BibitemOpen
  \bibfield  {author} {\bibinfo {author} {\bibfnamefont {S.}~\bibnamefont
  {{Borisenko}}}, \bibinfo {author} {\bibfnamefont {D.}~\bibnamefont
  {{Evtushinsky}}}, \bibinfo {author} {\bibfnamefont {Q.}~\bibnamefont
  {{Gibson}}}, \bibinfo {author} {\bibfnamefont {A.}~\bibnamefont {{Yaresko}}},
  \bibinfo {author} {\bibfnamefont {T.}~\bibnamefont {{Kim}}}, \bibinfo
  {author} {\bibfnamefont {M.~N.}\ \bibnamefont {{Ali}}}, \bibinfo {author}
  {\bibfnamefont {B.}~\bibnamefont {{Buechner}}}, \bibinfo {author}
  {\bibfnamefont {M.}~\bibnamefont {{Hoesch}}}, \ and\ \bibinfo {author}
  {\bibfnamefont {R.~J.}\ \bibnamefont {{Cava}}},\ }\href@noop {} {\ }\Eprint
  {http://arxiv.org/abs/1507.04847} {arXiv:1507.04847} \BibitemShut {NoStop}%
\bibitem [{\citenamefont {{Liu}}\ \emph {et~al.}()\citenamefont {{Liu}},
  \citenamefont {{Hu}}, \citenamefont {{Zhang}}, \citenamefont {{Graf}},
  \citenamefont {{Cao}}, \citenamefont {{Radmanesh}}, \citenamefont {{Adams}},
  \citenamefont {{Zhu}}, \citenamefont {{Cheng}}, \citenamefont {{Liu}},
  \citenamefont {{Phelan}}, \citenamefont {{Wei}}, \citenamefont {{Tennant}},
  \citenamefont {{DiTusa}}, \citenamefont {{Chiorescu}}, \citenamefont
  {{Spinu}},\ and\ \citenamefont {{Mao}}}]{weylexp9}%
  \BibitemOpen
  \bibfield  {author} {\bibinfo {author} {\bibfnamefont {J.~Y.}\ \bibnamefont
  {{Liu}}}, \bibinfo {author} {\bibfnamefont {J.}~\bibnamefont {{Hu}}},
  \bibinfo {author} {\bibfnamefont {Q.}~\bibnamefont {{Zhang}}}, \bibinfo
  {author} {\bibfnamefont {D.}~\bibnamefont {{Graf}}}, \bibinfo {author}
  {\bibfnamefont {H.~B.}\ \bibnamefont {{Cao}}}, \bibinfo {author}
  {\bibfnamefont {S.~M.~A.}\ \bibnamefont {{Radmanesh}}}, \bibinfo {author}
  {\bibfnamefont {D.~J.}\ \bibnamefont {{Adams}}}, \bibinfo {author}
  {\bibfnamefont {Y.~L.}\ \bibnamefont {{Zhu}}}, \bibinfo {author}
  {\bibfnamefont {G.~F.}\ \bibnamefont {{Cheng}}}, \bibinfo {author}
  {\bibfnamefont {X.}~\bibnamefont {{Liu}}}, \bibinfo {author} {\bibfnamefont
  {W.~A.}\ \bibnamefont {{Phelan}}}, \bibinfo {author} {\bibfnamefont
  {J.}~\bibnamefont {{Wei}}}, \bibinfo {author} {\bibfnamefont {D.~A.}\
  \bibnamefont {{Tennant}}}, \bibinfo {author} {\bibfnamefont {J.~F.}\
  \bibnamefont {{DiTusa}}}, \bibinfo {author} {\bibfnamefont {I.}~\bibnamefont
  {{Chiorescu}}}, \bibinfo {author} {\bibfnamefont {L.}~\bibnamefont
  {{Spinu}}}, \ and\ \bibinfo {author} {\bibfnamefont {Z.~Q.}\ \bibnamefont
  {{Mao}}},\ }\href@noop {} {\ }\Eprint {http://arxiv.org/abs/1507.07978}
  {arXiv:1507.07978} \BibitemShut {NoStop}%
\bibitem [{\citenamefont {Xu}\ \emph {et~al.}(2014)\citenamefont {Xu},
  \citenamefont {Chu},\ and\ \citenamefont {Zhang}}]{weylexp10}%
  \BibitemOpen
  \bibfield  {author} {\bibinfo {author} {\bibfnamefont {Y.}~\bibnamefont
  {Xu}}, \bibinfo {author} {\bibfnamefont {R.-L.}\ \bibnamefont {Chu}}, \ and\
  \bibinfo {author} {\bibfnamefont {C.}~\bibnamefont {Zhang}},\ }\href
  {\doibase 10.1103/PhysRevLett.112.136402} {\bibfield  {journal} {\bibinfo
  {journal} {Phys. Rev. Lett.}\ }\textbf {\bibinfo {volume} {112}},\ \bibinfo
  {pages} {136402} (\bibinfo {year} {2014})}\BibitemShut {NoStop}%
\bibitem [{\citenamefont {{Lu}}\ \emph {et~al.}(2015)\citenamefont {{Lu}},
  \citenamefont {{Wang}}, \citenamefont {{Ye}}, \citenamefont {{Ran}},
  \citenamefont {{Fu}}, \citenamefont {{Joannopoulos}},\ and\ \citenamefont
  {{Solja{\v c}i{\'c}}}}]{weylexp11}%
  \BibitemOpen
  \bibfield  {author} {\bibinfo {author} {\bibfnamefont {L.}~\bibnamefont
  {{Lu}}}, \bibinfo {author} {\bibfnamefont {Z.}~\bibnamefont {{Wang}}},
  \bibinfo {author} {\bibfnamefont {D.}~\bibnamefont {{Ye}}}, \bibinfo {author}
  {\bibfnamefont {L.}~\bibnamefont {{Ran}}}, \bibinfo {author} {\bibfnamefont
  {L.}~\bibnamefont {{Fu}}}, \bibinfo {author} {\bibfnamefont {J.~D.}\
  \bibnamefont {{Joannopoulos}}}, \ and\ \bibinfo {author} {\bibfnamefont
  {M.}~\bibnamefont {{Solja{\v c}i{\'c}}}},\ }\href {\doibase
  10.1126/science.aaa9273} {\bibfield  {journal} {\bibinfo  {journal}
  {Science}\ }\textbf {\bibinfo {volume} {349}},\ \bibinfo {pages} {622}
  (\bibinfo {year} {2015})}\BibitemShut {NoStop}%
\bibitem [{\citenamefont {{Xu}}\ \emph {et~al.}(2015)\citenamefont {{Xu}},
  \citenamefont {{Alidoust}}, \citenamefont {{Belopolski}}, \citenamefont
  {{Zhang}}, \citenamefont {{Bian}}, \citenamefont {{Chang}}, \citenamefont
  {{Zheng}}, \citenamefont {{Strokov}}, \citenamefont {{Sanchez}},
  \citenamefont {{Chang}}, \citenamefont {{Yuan}}, \citenamefont {{Mou}},
  \citenamefont {{Wu}}, \citenamefont {{Huang}}, \citenamefont {{Lee}},
  \citenamefont {{Huang}}, \citenamefont {{Wang}}, \citenamefont {{Bansil}},
  \citenamefont {{Jeng}}, \citenamefont {{Neupert}}, \citenamefont
  {{Kaminski}}, \citenamefont {{Lin}}, \citenamefont {{Jia}},\ and\
  \citenamefont {{Z. Hasan}}}]{weylexp12}%
  \BibitemOpen
  \bibfield  {author} {\bibinfo {author} {\bibfnamefont {S.-Y.}\ \bibnamefont
  {{Xu}}}, \bibinfo {author} {\bibfnamefont {N.}~\bibnamefont {{Alidoust}}},
  \bibinfo {author} {\bibfnamefont {I.}~\bibnamefont {{Belopolski}}}, \bibinfo
  {author} {\bibfnamefont {C.}~\bibnamefont {{Zhang}}}, \bibinfo {author}
  {\bibfnamefont {G.}~\bibnamefont {{Bian}}}, \bibinfo {author} {\bibfnamefont
  {T.-R.}\ \bibnamefont {{Chang}}}, \bibinfo {author} {\bibfnamefont
  {H.}~\bibnamefont {{Zheng}}}, \bibinfo {author} {\bibfnamefont
  {V.}~\bibnamefont {{Strokov}}}, \bibinfo {author} {\bibfnamefont {D.~S.}\
  \bibnamefont {{Sanchez}}}, \bibinfo {author} {\bibfnamefont {G.}~\bibnamefont
  {{Chang}}}, \bibinfo {author} {\bibfnamefont {Z.}~\bibnamefont {{Yuan}}},
  \bibinfo {author} {\bibfnamefont {D.}~\bibnamefont {{Mou}}}, \bibinfo
  {author} {\bibfnamefont {Y.}~\bibnamefont {{Wu}}}, \bibinfo {author}
  {\bibfnamefont {L.}~\bibnamefont {{Huang}}}, \bibinfo {author} {\bibfnamefont
  {C.-C.}\ \bibnamefont {{Lee}}}, \bibinfo {author} {\bibfnamefont {S.-M.}\
  \bibnamefont {{Huang}}}, \bibinfo {author} {\bibfnamefont {B.}~\bibnamefont
  {{Wang}}}, \bibinfo {author} {\bibfnamefont {A.}~\bibnamefont {{Bansil}}},
  \bibinfo {author} {\bibfnamefont {H.-T.}\ \bibnamefont {{Jeng}}}, \bibinfo
  {author} {\bibfnamefont {T.}~\bibnamefont {{Neupert}}}, \bibinfo {author}
  {\bibfnamefont {A.}~\bibnamefont {{Kaminski}}}, \bibinfo {author}
  {\bibfnamefont {H.}~\bibnamefont {{Lin}}}, \bibinfo {author} {\bibfnamefont
  {S.}~\bibnamefont {{Jia}}}, \ and\ \bibinfo {author} {\bibfnamefont
  {M.}~\bibnamefont {{Z. Hasan}}},\ }\href
  {http://dx.doi.org/10.1038/nphys3437} {\bibfield  {journal} {\bibinfo
  {journal} {Nat. Phys.}\ }\textbf {\bibinfo {volume} {11}},\ \bibinfo {pages}
  {748} (\bibinfo {year} {2015})}\BibitemShut {NoStop}%
\bibitem [{\citenamefont {{Sushkov}}\ \emph {et~al.}(2015)\citenamefont
  {{Sushkov}}, \citenamefont {{Hofmann}}, \citenamefont {{Jenkins}},
  \citenamefont {{Ishikawa}}, \citenamefont {{Nakatsuji}}, \citenamefont {{Das
  Sarma}},\ and\ \citenamefont {{Drew}}}]{weylexp13}%
  \BibitemOpen
  \bibfield  {author} {\bibinfo {author} {\bibfnamefont {A.~B.}\ \bibnamefont
  {{Sushkov}}}, \bibinfo {author} {\bibfnamefont {J.~B.}\ \bibnamefont
  {{Hofmann}}}, \bibinfo {author} {\bibfnamefont {G.~S.}\ \bibnamefont
  {{Jenkins}}}, \bibinfo {author} {\bibfnamefont {J.}~\bibnamefont
  {{Ishikawa}}}, \bibinfo {author} {\bibfnamefont {S.}~\bibnamefont
  {{Nakatsuji}}}, \bibinfo {author} {\bibfnamefont {S.}~\bibnamefont {{Das
  Sarma}}}, \ and\ \bibinfo {author} {\bibfnamefont {H.~D.}\ \bibnamefont
  {{Drew}}},\ }\href {\doibase 10.1103/PhysRevB.92.241108} {\bibfield
  {journal} {\bibinfo  {journal} {Phys. Rev. B}\ }\textbf {\bibinfo {volume}
  {92}},\ \bibinfo {eid} {241108} (\bibinfo {year} {2015})}\BibitemShut
  {NoStop}%
\bibitem [{\citenamefont {{Zhang}}\ \emph {et~al.}(2016)\citenamefont
  {{Zhang}}, \citenamefont {{Xu}}, \citenamefont {{Belopolski}}, \citenamefont
  {{Yuan}}, \citenamefont {{Lin}}, \citenamefont {{Tong}}, \citenamefont
  {{Alidoust}}, \citenamefont {{Lee}}, \citenamefont {{Huang}}, \citenamefont
  {{Lin}}, \citenamefont {{Neupane}}, \citenamefont {{Sanchez}}, \citenamefont
  {{Zheng}}, \citenamefont {{Bian}}, \citenamefont {{Wang}}, \citenamefont
  {{Zhang}}, \citenamefont {{Neupert}}, \citenamefont {{Z. Hasan}},\ and\
  \citenamefont {{Jia}}}]{weylexp14}%
  \BibitemOpen
  \bibfield  {author} {\bibinfo {author} {\bibfnamefont {C.}~\bibnamefont
  {{Zhang}}}, \bibinfo {author} {\bibfnamefont {S.-Y.}\ \bibnamefont {{Xu}}},
  \bibinfo {author} {\bibfnamefont {I.}~\bibnamefont {{Belopolski}}}, \bibinfo
  {author} {\bibfnamefont {Z.}~\bibnamefont {{Yuan}}}, \bibinfo {author}
  {\bibfnamefont {Z.}~\bibnamefont {{Lin}}}, \bibinfo {author} {\bibfnamefont
  {B.}~\bibnamefont {{Tong}}}, \bibinfo {author} {\bibfnamefont
  {N.}~\bibnamefont {{Alidoust}}}, \bibinfo {author} {\bibfnamefont {C.-C.}\
  \bibnamefont {{Lee}}}, \bibinfo {author} {\bibfnamefont {S.-M.}\ \bibnamefont
  {{Huang}}}, \bibinfo {author} {\bibfnamefont {H.}~\bibnamefont {{Lin}}},
  \bibinfo {author} {\bibfnamefont {M.}~\bibnamefont {{Neupane}}}, \bibinfo
  {author} {\bibfnamefont {D.~S.}\ \bibnamefont {{Sanchez}}}, \bibinfo {author}
  {\bibfnamefont {H.}~\bibnamefont {{Zheng}}}, \bibinfo {author} {\bibfnamefont
  {G.}~\bibnamefont {{Bian}}}, \bibinfo {author} {\bibfnamefont
  {J.}~\bibnamefont {{Wang}}}, \bibinfo {author} {\bibfnamefont
  {C.}~\bibnamefont {{Zhang}}}, \bibinfo {author} {\bibfnamefont
  {T.}~\bibnamefont {{Neupert}}}, \bibinfo {author} {\bibfnamefont
  {M.}~\bibnamefont {{Z. Hasan}}}, \ and\ \bibinfo {author} {\bibfnamefont
  {S.}~\bibnamefont {{Jia}}},\ }\href {http://dx.doi.org/10.1038/ncomms10735}
  {\bibfield  {journal} {\bibinfo  {journal} {Nat. Commun.}\ }\textbf {\bibinfo
  {volume} {7}},\ \bibinfo {pages} {10735} (\bibinfo {year}
  {2016})}\BibitemShut {NoStop}%
\bibitem [{\citenamefont {Du}\ \emph {et~al.}(2016)\citenamefont {Du},
  \citenamefont {Wang}, \citenamefont {Chen}, \citenamefont {Mao},
  \citenamefont {Khan}, \citenamefont {Xu}, \citenamefont {Zhou}, \citenamefont
  {Zhang}, \citenamefont {Yang}, \citenamefont {Chen}, \citenamefont {Feng},\
  and\ \citenamefont {Fang}}]{weylexp15}%
  \BibitemOpen
  \bibfield  {author} {\bibinfo {author} {\bibfnamefont {J.}~\bibnamefont
  {Du}}, \bibinfo {author} {\bibfnamefont {H.}~\bibnamefont {Wang}}, \bibinfo
  {author} {\bibfnamefont {Q.}~\bibnamefont {Chen}}, \bibinfo {author}
  {\bibfnamefont {Q.}~\bibnamefont {Mao}}, \bibinfo {author} {\bibfnamefont
  {R.}~\bibnamefont {Khan}}, \bibinfo {author} {\bibfnamefont {B.}~\bibnamefont
  {Xu}}, \bibinfo {author} {\bibfnamefont {Y.}~\bibnamefont {Zhou}}, \bibinfo
  {author} {\bibfnamefont {Y.}~\bibnamefont {Zhang}}, \bibinfo {author}
  {\bibfnamefont {J.}~\bibnamefont {Yang}}, \bibinfo {author} {\bibfnamefont
  {B.}~\bibnamefont {Chen}}, \bibinfo {author} {\bibfnamefont {C.}~\bibnamefont
  {Feng}}, \ and\ \bibinfo {author} {\bibfnamefont {M.}~\bibnamefont {Fang}},\
  }\href {\doibase 10.1007/s11433-016-5798-4} {\bibfield  {journal} {\bibinfo
  {journal} {Sci. China. Phys. Mech.}\ }\textbf {\bibinfo {volume} {59}},\
  \bibinfo {pages} {657406} (\bibinfo {year} {2016})}\BibitemShut {NoStop}%
\bibitem [{\citenamefont {{Arnold}}\ \emph {et~al.}(2016)\citenamefont
  {{Arnold}}, \citenamefont {{Shekhar}}, \citenamefont {{Wu}}, \citenamefont
  {{Sun}}, \citenamefont {{Dos Reis}}, \citenamefont {{Kumar}}, \citenamefont
  {{Naumann}}, \citenamefont {{Ajeesh}}, \citenamefont {{Schmidt}},
  \citenamefont {{Grushin}}, \citenamefont {{Bardarson}}, \citenamefont
  {{Baenitz}}, \citenamefont {{Sokolov}}, \citenamefont {{Borrmann}},
  \citenamefont {{Nicklas}}, \citenamefont {{Felser}}, \citenamefont
  {{Hassinger}},\ and\ \citenamefont {{Yan}}}]{weylexp16}%
  \BibitemOpen
  \bibfield  {author} {\bibinfo {author} {\bibfnamefont {F.}~\bibnamefont
  {{Arnold}}}, \bibinfo {author} {\bibfnamefont {C.}~\bibnamefont {{Shekhar}}},
  \bibinfo {author} {\bibfnamefont {S.-C.}\ \bibnamefont {{Wu}}}, \bibinfo
  {author} {\bibfnamefont {Y.}~\bibnamefont {{Sun}}}, \bibinfo {author}
  {\bibfnamefont {R.~D.}\ \bibnamefont {{Dos Reis}}}, \bibinfo {author}
  {\bibfnamefont {N.}~\bibnamefont {{Kumar}}}, \bibinfo {author} {\bibfnamefont
  {M.}~\bibnamefont {{Naumann}}}, \bibinfo {author} {\bibfnamefont {M.~O.}\
  \bibnamefont {{Ajeesh}}}, \bibinfo {author} {\bibfnamefont {M.}~\bibnamefont
  {{Schmidt}}}, \bibinfo {author} {\bibfnamefont {A.~G.}\ \bibnamefont
  {{Grushin}}}, \bibinfo {author} {\bibfnamefont {J.~H.}\ \bibnamefont
  {{Bardarson}}}, \bibinfo {author} {\bibfnamefont {M.}~\bibnamefont
  {{Baenitz}}}, \bibinfo {author} {\bibfnamefont {D.}~\bibnamefont
  {{Sokolov}}}, \bibinfo {author} {\bibfnamefont {H.}~\bibnamefont
  {{Borrmann}}}, \bibinfo {author} {\bibfnamefont {M.}~\bibnamefont
  {{Nicklas}}}, \bibinfo {author} {\bibfnamefont {C.}~\bibnamefont {{Felser}}},
  \bibinfo {author} {\bibfnamefont {E.}~\bibnamefont {{Hassinger}}}, \ and\
  \bibinfo {author} {\bibfnamefont {B.}~\bibnamefont {{Yan}}},\ }\href
  {\doibase 10.1038/ncomms11615} {\bibfield  {journal} {\bibinfo  {journal}
  {Nat. Commun.}\ }\textbf {\bibinfo {volume} {7}},\ \bibinfo {eid} {11615}
  (\bibinfo {year} {2016})}\BibitemShut {NoStop}%
\bibitem [{\citenamefont {{Huang}}\ \emph {et~al.}(2015)\citenamefont
  {{Huang}}, \citenamefont {{Zhao}}, \citenamefont {{Long}}, \citenamefont
  {{Wang}}, \citenamefont {{Chen}}, \citenamefont {{Yang}}, \citenamefont
  {{Liang}}, \citenamefont {{Xue}}, \citenamefont {{Weng}}, \citenamefont
  {{Fang}}, \citenamefont {{Dai}},\ and\ \citenamefont {{Chen}}}]{weylexp17}%
  \BibitemOpen
  \bibfield  {author} {\bibinfo {author} {\bibfnamefont {X.}~\bibnamefont
  {{Huang}}}, \bibinfo {author} {\bibfnamefont {L.}~\bibnamefont {{Zhao}}},
  \bibinfo {author} {\bibfnamefont {Y.}~\bibnamefont {{Long}}}, \bibinfo
  {author} {\bibfnamefont {P.}~\bibnamefont {{Wang}}}, \bibinfo {author}
  {\bibfnamefont {D.}~\bibnamefont {{Chen}}}, \bibinfo {author} {\bibfnamefont
  {Z.}~\bibnamefont {{Yang}}}, \bibinfo {author} {\bibfnamefont
  {H.}~\bibnamefont {{Liang}}}, \bibinfo {author} {\bibfnamefont
  {M.}~\bibnamefont {{Xue}}}, \bibinfo {author} {\bibfnamefont
  {H.}~\bibnamefont {{Weng}}}, \bibinfo {author} {\bibfnamefont
  {Z.}~\bibnamefont {{Fang}}}, \bibinfo {author} {\bibfnamefont
  {X.}~\bibnamefont {{Dai}}}, \ and\ \bibinfo {author} {\bibfnamefont
  {G.}~\bibnamefont {{Chen}}},\ }\href {\doibase 10.1103/PhysRevX.5.031023}
  {\bibfield  {journal} {\bibinfo  {journal} {Phys. Rev. X}\ }\textbf {\bibinfo
  {volume} {5}},\ \bibinfo {eid} {031023} (\bibinfo {year} {2015})}\BibitemShut
  {NoStop}%
\bibitem [{\citenamefont {{Wang}}\ \emph {et~al.}(2016)\citenamefont {{Wang}},
  \citenamefont {{Zheng}}, \citenamefont {{Shen}}, \citenamefont {{Lu}},
  \citenamefont {{Fang}}, \citenamefont {{Sheng}}, \citenamefont {{Zhou}},
  \citenamefont {{Yang}}, \citenamefont {{Li}}, \citenamefont {{Feng}},\ and\
  \citenamefont {{Xu}}}]{weylexp18}%
  \BibitemOpen
  \bibfield  {author} {\bibinfo {author} {\bibfnamefont {Z.}~\bibnamefont
  {{Wang}}}, \bibinfo {author} {\bibfnamefont {Y.}~\bibnamefont {{Zheng}}},
  \bibinfo {author} {\bibfnamefont {Z.}~\bibnamefont {{Shen}}}, \bibinfo
  {author} {\bibfnamefont {Y.}~\bibnamefont {{Lu}}}, \bibinfo {author}
  {\bibfnamefont {H.}~\bibnamefont {{Fang}}}, \bibinfo {author} {\bibfnamefont
  {F.}~\bibnamefont {{Sheng}}}, \bibinfo {author} {\bibfnamefont
  {Y.}~\bibnamefont {{Zhou}}}, \bibinfo {author} {\bibfnamefont
  {X.}~\bibnamefont {{Yang}}}, \bibinfo {author} {\bibfnamefont
  {Y.}~\bibnamefont {{Li}}}, \bibinfo {author} {\bibfnamefont {C.}~\bibnamefont
  {{Feng}}}, \ and\ \bibinfo {author} {\bibfnamefont {Z.-A.}\ \bibnamefont
  {{Xu}}},\ }\href {\doibase 10.1103/PhysRevB.93.121112} {\bibfield  {journal}
  {\bibinfo  {journal} {Phys. Rev. B}\ }\textbf {\bibinfo {volume} {93}},\
  \bibinfo {eid} {121112} (\bibinfo {year} {2016})}\BibitemShut {NoStop}%
\bibitem [{\citenamefont {{Bednik}}\ \emph {et~al.}(2015)\citenamefont
  {{Bednik}}, \citenamefont {{Zyuzin}},\ and\ \citenamefont
  {{Burkov}}}]{weylsc1}%
  \BibitemOpen
  \bibfield  {author} {\bibinfo {author} {\bibfnamefont {G.}~\bibnamefont
  {{Bednik}}}, \bibinfo {author} {\bibfnamefont {A.~A.}\ \bibnamefont
  {{Zyuzin}}}, \ and\ \bibinfo {author} {\bibfnamefont {A.~A.}\ \bibnamefont
  {{Burkov}}},\ }\href {\doibase 10.1103/PhysRevB.92.035153} {\bibfield
  {journal} {\bibinfo  {journal} {Phys. Rev. B}\ }\textbf {\bibinfo {volume}
  {92}},\ \bibinfo {eid} {035153} (\bibinfo {year} {2015})}\BibitemShut
  {NoStop}%
\bibitem [{\citenamefont {{Li}}\ and\ \citenamefont {{Haldane}}()}]{weylsc2}%
  \BibitemOpen
  \bibfield  {author} {\bibinfo {author} {\bibfnamefont {Y.}~\bibnamefont
  {{Li}}}\ and\ \bibinfo {author} {\bibfnamefont {F.~D.~M.}\ \bibnamefont
  {{Haldane}}},\ }\href@noop {} {\ }\Eprint {http://arxiv.org/abs/1510.01730}
  {arXiv:1510.01730} \BibitemShut {NoStop}%
\bibitem [{\citenamefont {{Zhou}}\ \emph {et~al.}(2016)\citenamefont {{Zhou}},
  \citenamefont {{Gao}},\ and\ \citenamefont {{Wang}}}]{weylsc3}%
  \BibitemOpen
  \bibfield  {author} {\bibinfo {author} {\bibfnamefont {T.}~\bibnamefont
  {{Zhou}}}, \bibinfo {author} {\bibfnamefont {Y.}~\bibnamefont {{Gao}}}, \
  and\ \bibinfo {author} {\bibfnamefont {Z.~D.}\ \bibnamefont {{Wang}}},\
  }\href {\doibase 10.1103/PhysRevB.93.094517} {\bibfield  {journal} {\bibinfo
  {journal} {Phys. Rev. B}\ }\textbf {\bibinfo {volume} {93}},\ \bibinfo {eid}
  {094517} (\bibinfo {year} {2016})}\BibitemShut {NoStop}%
\bibitem [{\citenamefont {{Cho}}\ \emph {et~al.}(2012)\citenamefont {{Cho}},
  \citenamefont {{Bardarson}}, \citenamefont {{Lu}},\ and\ \citenamefont
  {{Moore}}}]{weylsc4}%
  \BibitemOpen
  \bibfield  {author} {\bibinfo {author} {\bibfnamefont {G.~Y.}\ \bibnamefont
  {{Cho}}}, \bibinfo {author} {\bibfnamefont {J.~H.}\ \bibnamefont
  {{Bardarson}}}, \bibinfo {author} {\bibfnamefont {Y.-M.}\ \bibnamefont
  {{Lu}}}, \ and\ \bibinfo {author} {\bibfnamefont {J.~E.}\ \bibnamefont
  {{Moore}}},\ }\href {\doibase 10.1103/PhysRevB.86.214514} {\bibfield
  {journal} {\bibinfo  {journal} {Phys. Rev. B}\ }\textbf {\bibinfo {volume}
  {86}},\ \bibinfo {eid} {214514} (\bibinfo {year} {2012})}\BibitemShut
  {NoStop}%
\bibitem [{\citenamefont {Wei}\ \emph {et~al.}(2014)\citenamefont {Wei},
  \citenamefont {Chao},\ and\ \citenamefont {Aji}}]{weylsc5}%
  \BibitemOpen
  \bibfield  {author} {\bibinfo {author} {\bibfnamefont {H.}~\bibnamefont
  {Wei}}, \bibinfo {author} {\bibfnamefont {S.-P.}\ \bibnamefont {Chao}}, \
  and\ \bibinfo {author} {\bibfnamefont {V.}~\bibnamefont {Aji}},\ }\href
  {\doibase 10.1103/PhysRevB.89.014506} {\bibfield  {journal} {\bibinfo
  {journal} {Phys. Rev. B}\ }\textbf {\bibinfo {volume} {89}},\ \bibinfo
  {pages} {014506} (\bibinfo {year} {2014})}\BibitemShut {NoStop}%
\bibitem [{\citenamefont {Meng}\ and\ \citenamefont
  {Balents}(2012)}]{Meng_PRB_86_054504}%
  \BibitemOpen
  \bibfield  {author} {\bibinfo {author} {\bibfnamefont {T.}~\bibnamefont
  {Meng}}\ and\ \bibinfo {author} {\bibfnamefont {L.}~\bibnamefont {Balents}},\
  }\href {\doibase 10.1103/PhysRevB.86.054504} {\bibfield  {journal} {\bibinfo
  {journal} {Phys. Rev. B}\ }\textbf {\bibinfo {volume} {86}},\ \bibinfo
  {pages} {054504} (\bibinfo {year} {2012})}\BibitemShut {NoStop}%
\bibitem [{\citenamefont {Lu}\ \emph {et~al.}(2015)\citenamefont {Lu},
  \citenamefont {Yada}, \citenamefont {Sato},\ and\ \citenamefont
  {Tanaka}}]{Lu_PRL_114_096804}%
  \BibitemOpen
  \bibfield  {author} {\bibinfo {author} {\bibfnamefont {B.}~\bibnamefont
  {Lu}}, \bibinfo {author} {\bibfnamefont {K.}~\bibnamefont {Yada}}, \bibinfo
  {author} {\bibfnamefont {M.}~\bibnamefont {Sato}}, \ and\ \bibinfo {author}
  {\bibfnamefont {Y.}~\bibnamefont {Tanaka}},\ }\href {\doibase
  10.1103/PhysRevLett.114.096804} {\bibfield  {journal} {\bibinfo  {journal}
  {Phys. Rev. Lett.}\ }\textbf {\bibinfo {volume} {114}},\ \bibinfo {pages}
  {096804} (\bibinfo {year} {2015})}\BibitemShut {NoStop}%
\bibitem [{\citenamefont {{Qi}}\ \emph {et~al.}(2016)\citenamefont {{Qi}},
  \citenamefont {{Naumov}}, \citenamefont {{Ali}}, \citenamefont {{Rajamathi}},
  \citenamefont {{Schnelle}}, \citenamefont {{Barkalov}}, \citenamefont
  {{Hanfland}}, \citenamefont {{Wu}}, \citenamefont {{Shekhar}}, \citenamefont
  {{Sun}}, \citenamefont {{S{\"u}{\ss}}}, \citenamefont {{Schmidt}},
  \citenamefont {{Schwarz}}, \citenamefont {{Pippel}}, \citenamefont
  {{Werner}}, \citenamefont {{Hillebrand}}, \citenamefont {{F{\"o}rster}},
  \citenamefont {{Kampert}}, \citenamefont {{Parkin}}, \citenamefont {{Cava}},
  \citenamefont {{Felser}}, \citenamefont {{Yan}},\ and\ \citenamefont
  {{Medvedev}}}]{weylscexp}%
  \BibitemOpen
  \bibfield  {author} {\bibinfo {author} {\bibfnamefont {Y.}~\bibnamefont
  {{Qi}}}, \bibinfo {author} {\bibfnamefont {P.~G.}\ \bibnamefont {{Naumov}}},
  \bibinfo {author} {\bibfnamefont {M.~N.}\ \bibnamefont {{Ali}}}, \bibinfo
  {author} {\bibfnamefont {C.~R.}\ \bibnamefont {{Rajamathi}}}, \bibinfo
  {author} {\bibfnamefont {W.}~\bibnamefont {{Schnelle}}}, \bibinfo {author}
  {\bibfnamefont {O.}~\bibnamefont {{Barkalov}}}, \bibinfo {author}
  {\bibfnamefont {M.}~\bibnamefont {{Hanfland}}}, \bibinfo {author}
  {\bibfnamefont {S.-C.}\ \bibnamefont {{Wu}}}, \bibinfo {author}
  {\bibfnamefont {C.}~\bibnamefont {{Shekhar}}}, \bibinfo {author}
  {\bibfnamefont {Y.}~\bibnamefont {{Sun}}}, \bibinfo {author} {\bibfnamefont
  {V.}~\bibnamefont {{S{\"u}{\ss}}}}, \bibinfo {author} {\bibfnamefont
  {M.}~\bibnamefont {{Schmidt}}}, \bibinfo {author} {\bibfnamefont
  {U.}~\bibnamefont {{Schwarz}}}, \bibinfo {author} {\bibfnamefont
  {E.}~\bibnamefont {{Pippel}}}, \bibinfo {author} {\bibfnamefont
  {P.}~\bibnamefont {{Werner}}}, \bibinfo {author} {\bibfnamefont
  {R.}~\bibnamefont {{Hillebrand}}}, \bibinfo {author} {\bibfnamefont
  {T.}~\bibnamefont {{F{\"o}rster}}}, \bibinfo {author} {\bibfnamefont
  {E.}~\bibnamefont {{Kampert}}}, \bibinfo {author} {\bibfnamefont
  {S.}~\bibnamefont {{Parkin}}}, \bibinfo {author} {\bibfnamefont {R.~J.}\
  \bibnamefont {{Cava}}}, \bibinfo {author} {\bibfnamefont {C.}~\bibnamefont
  {{Felser}}}, \bibinfo {author} {\bibfnamefont {B.}~\bibnamefont {{Yan}}}, \
  and\ \bibinfo {author} {\bibfnamefont {S.~A.}\ \bibnamefont {{Medvedev}}},\
  }\href {\doibase 10.1038/ncomms11038} {\bibfield  {journal} {\bibinfo
  {journal} {Nat. Commun.}\ }\textbf {\bibinfo {volume} {7}},\ \bibinfo {eid}
  {11038} (\bibinfo {year} {2016})}\BibitemShut {NoStop}%
\bibitem [{\citenamefont {{Li}}\ \emph {et~al.}()\citenamefont {{Li}},
  \citenamefont {{Zhou}}, \citenamefont {{Guo}}, \citenamefont {{Chen}},
  \citenamefont {{Lu}}, \citenamefont {{Wang}}, \citenamefont {{An}},
  \citenamefont {{Zhou}}, \citenamefont {{Xing}}, \citenamefont {{Du}},
  \citenamefont {{Zhu}}, \citenamefont {{Yang}}, \citenamefont {{Sun}},
  \citenamefont {{Yang}}, \citenamefont {{Zhang}},\ and\ \citenamefont
  {{Wen}}}]{arXiv_1611_02548}%
  \BibitemOpen
  \bibfield  {author} {\bibinfo {author} {\bibfnamefont {Y.}~\bibnamefont
  {{Li}}}, \bibinfo {author} {\bibfnamefont {Y.}~\bibnamefont {{Zhou}}},
  \bibinfo {author} {\bibfnamefont {Z.}~\bibnamefont {{Guo}}}, \bibinfo
  {author} {\bibfnamefont {X.}~\bibnamefont {{Chen}}}, \bibinfo {author}
  {\bibfnamefont {P.}~\bibnamefont {{Lu}}}, \bibinfo {author} {\bibfnamefont
  {X.}~\bibnamefont {{Wang}}}, \bibinfo {author} {\bibfnamefont
  {C.}~\bibnamefont {{An}}}, \bibinfo {author} {\bibfnamefont {Y.}~\bibnamefont
  {{Zhou}}}, \bibinfo {author} {\bibfnamefont {J.}~\bibnamefont {{Xing}}},
  \bibinfo {author} {\bibfnamefont {G.}~\bibnamefont {{Du}}}, \bibinfo {author}
  {\bibfnamefont {X.}~\bibnamefont {{Zhu}}}, \bibinfo {author} {\bibfnamefont
  {H.}~\bibnamefont {{Yang}}}, \bibinfo {author} {\bibfnamefont
  {J.}~\bibnamefont {{Sun}}}, \bibinfo {author} {\bibfnamefont
  {Z.}~\bibnamefont {{Yang}}}, \bibinfo {author} {\bibfnamefont
  {Y.}~\bibnamefont {{Zhang}}}, \ and\ \bibinfo {author} {\bibfnamefont
  {H.-H.}\ \bibnamefont {{Wen}}},\ }\href@noop {} {\ }\Eprint
  {http://arxiv.org/abs/1611.02548} {arXiv:1611.02548} \BibitemShut {NoStop}%
\bibitem [{\citenamefont {{Khanna}}\ \emph {et~al.}(2016)\citenamefont
  {{Khanna}}, \citenamefont {{Mukherjee}}, \citenamefont {{Kundu}},\ and\
  \citenamefont {{Rao}}}]{josephsonweylcompare}%
  \BibitemOpen
  \bibfield  {author} {\bibinfo {author} {\bibfnamefont {U.}~\bibnamefont
  {{Khanna}}}, \bibinfo {author} {\bibfnamefont {D.~K.}\ \bibnamefont
  {{Mukherjee}}}, \bibinfo {author} {\bibfnamefont {A.}~\bibnamefont
  {{Kundu}}}, \ and\ \bibinfo {author} {\bibfnamefont {S.}~\bibnamefont
  {{Rao}}},\ }\href {\doibase 10.1103/PhysRevB.93.121409} {\bibfield  {journal}
  {\bibinfo  {journal} {Phys. Rev. B}\ }\textbf {\bibinfo {volume} {93}},\
  \bibinfo {eid} {121409} (\bibinfo {year} {2016})}\BibitemShut {NoStop}%
\bibitem [{\citenamefont {Kim}\ \emph {et~al.}(2016)\citenamefont {Kim},
  \citenamefont {Park},\ and\ \citenamefont {Gilbert}}]{PRB_93_214511}%
  \BibitemOpen
  \bibfield  {author} {\bibinfo {author} {\bibfnamefont {Y.}~\bibnamefont
  {Kim}}, \bibinfo {author} {\bibfnamefont {M.~J.}\ \bibnamefont {Park}}, \
  and\ \bibinfo {author} {\bibfnamefont {M.~J.}\ \bibnamefont {Gilbert}},\
  }\href {\doibase 10.1103/PhysRevB.93.214511} {\bibfield  {journal} {\bibinfo
  {journal} {Phys. Rev. B}\ }\textbf {\bibinfo {volume} {93}},\ \bibinfo
  {pages} {214511} (\bibinfo {year} {2016})}\BibitemShut {NoStop}%
\bibitem [{\citenamefont {{Heersche}}\ \emph {et~al.}(2007)\citenamefont
  {{Heersche}}, \citenamefont {{Jarillo-Herrero}}, \citenamefont {{Oostinga}},
  \citenamefont {{Vandersypen}},\ and\ \citenamefont
  {{Morpurgo}}}]{Heersche_Nature_446_56_2007}%
  \BibitemOpen
  \bibfield  {author} {\bibinfo {author} {\bibfnamefont {H.~B.}\ \bibnamefont
  {{Heersche}}}, \bibinfo {author} {\bibfnamefont {P.}~\bibnamefont
  {{Jarillo-Herrero}}}, \bibinfo {author} {\bibfnamefont {J.~B.}\ \bibnamefont
  {{Oostinga}}}, \bibinfo {author} {\bibfnamefont {L.~M.~K.}\ \bibnamefont
  {{Vandersypen}}}, \ and\ \bibinfo {author} {\bibfnamefont {A.~F.}\
  \bibnamefont {{Morpurgo}}},\ }\href {\doibase 10.1038/nature05555} {\bibfield
   {journal} {\bibinfo  {journal} {Nature (London)}\ }\textbf {\bibinfo
  {volume} {446}},\ \bibinfo {pages} {56} (\bibinfo {year} {2007})}\BibitemShut
  {NoStop}%
\bibitem [{\citenamefont {Heersche}\ \emph {et~al.}(2007)\citenamefont
  {Heersche}, \citenamefont {Jarillo-Herrero}, \citenamefont {Oostinga},
  \citenamefont {Vandersypen},\ and\ \citenamefont
  {Morpurgo}}]{Heersche_Solid_State_Commun_143_72_2007}%
  \BibitemOpen
  \bibfield  {author} {\bibinfo {author} {\bibfnamefont {H.~B.}\ \bibnamefont
  {Heersche}}, \bibinfo {author} {\bibfnamefont {P.}~\bibnamefont
  {Jarillo-Herrero}}, \bibinfo {author} {\bibfnamefont {J.~B.}\ \bibnamefont
  {Oostinga}}, \bibinfo {author} {\bibfnamefont {L.~M.}\ \bibnamefont
  {Vandersypen}}, \ and\ \bibinfo {author} {\bibfnamefont {A.~F.}\ \bibnamefont
  {Morpurgo}},\ }\href {\doibase http://dx.doi.org/10.1016/j.ssc.2007.02.044}
  {\bibfield  {journal} {\bibinfo  {journal} {Solid State Commun.}\ }\textbf
  {\bibinfo {volume} {143}},\ \bibinfo {pages} {72 } (\bibinfo {year}
  {2007})}\BibitemShut {NoStop}%
\bibitem [{\citenamefont {Takane}\ and\ \citenamefont
  {Imura}(2012)}]{Kanda_Physica_C_470_1477_2010}%
  \BibitemOpen
  \bibfield  {author} {\bibinfo {author} {\bibfnamefont {Y.}~\bibnamefont
  {Takane}}\ and\ \bibinfo {author} {\bibfnamefont {K.-I.}\ \bibnamefont
  {Imura}},\ }\href {\doibase 10.1143/JPSJ.81.094707} {\bibfield  {journal}
  {\bibinfo  {journal} {J. Phys. Soc. Japan}\ }\textbf {\bibinfo {volume}
  {81}},\ \bibinfo {pages} {094707} (\bibinfo {year} {2012})}\BibitemShut
  {NoStop}%
\bibitem [{\citenamefont {English}\ \emph {et~al.}(2016)\citenamefont
  {English}, \citenamefont {Hamilton}, \citenamefont {Chialvo}, \citenamefont
  {Moraru}, \citenamefont {Mason},\ and\ \citenamefont
  {Van~Harlingen}}]{English_PRB_94_115435_2016}%
  \BibitemOpen
  \bibfield  {author} {\bibinfo {author} {\bibfnamefont {C.~D.}\ \bibnamefont
  {English}}, \bibinfo {author} {\bibfnamefont {D.~R.}\ \bibnamefont
  {Hamilton}}, \bibinfo {author} {\bibfnamefont {C.}~\bibnamefont {Chialvo}},
  \bibinfo {author} {\bibfnamefont {I.~C.}\ \bibnamefont {Moraru}}, \bibinfo
  {author} {\bibfnamefont {N.}~\bibnamefont {Mason}}, \ and\ \bibinfo {author}
  {\bibfnamefont {D.~J.}\ \bibnamefont {Van~Harlingen}},\ }\href {\doibase
  10.1103/PhysRevB.94.115435} {\bibfield  {journal} {\bibinfo  {journal} {Phys.
  Rev. B}\ }\textbf {\bibinfo {volume} {94}},\ \bibinfo {pages} {115435}
  (\bibinfo {year} {2016})}\BibitemShut {NoStop}%
\bibitem [{\citenamefont {Kurter}\ \emph {et~al.}(2015)\citenamefont {Kurter},
  \citenamefont {Finck}, \citenamefont {Hor},\ and\ \citenamefont
  {Van~Harlingen}}]{Kurter_Nat_Commun_6_7130_2015}%
  \BibitemOpen
  \bibfield  {author} {\bibinfo {author} {\bibfnamefont {C.}~\bibnamefont
  {Kurter}}, \bibinfo {author} {\bibfnamefont {A.}~\bibnamefont {Finck}},
  \bibinfo {author} {\bibfnamefont {Y.~S.}\ \bibnamefont {Hor}}, \ and\
  \bibinfo {author} {\bibfnamefont {D.~J.}\ \bibnamefont {Van~Harlingen}},\
  }\href {http://dx.doi.org/10.1038/ncomms8130} {\bibfield  {journal} {\bibinfo
   {journal} {Nat. Commun.}\ }\textbf {\bibinfo {volume} {6}},\ \bibinfo
  {pages} {7130} (\bibinfo {year} {2015})}\BibitemShut {NoStop}%
\bibitem [{\citenamefont {Komatsu}\ \emph {et~al.}(2012)\citenamefont
  {Komatsu}, \citenamefont {Li}, \citenamefont {Autier-Laurent}, \citenamefont
  {Bouchiat},\ and\ \citenamefont {Gu\'eron}}]{Komatsu_PRB_86_115412_2012}%
  \BibitemOpen
  \bibfield  {author} {\bibinfo {author} {\bibfnamefont {K.}~\bibnamefont
  {Komatsu}}, \bibinfo {author} {\bibfnamefont {C.}~\bibnamefont {Li}},
  \bibinfo {author} {\bibfnamefont {S.}~\bibnamefont {Autier-Laurent}},
  \bibinfo {author} {\bibfnamefont {H.}~\bibnamefont {Bouchiat}}, \ and\
  \bibinfo {author} {\bibfnamefont {S.}~\bibnamefont {Gu\'eron}},\ }\href
  {\doibase 10.1103/PhysRevB.86.115412} {\bibfield  {journal} {\bibinfo
  {journal} {Phys. Rev. B}\ }\textbf {\bibinfo {volume} {86}},\ \bibinfo
  {pages} {115412} (\bibinfo {year} {2012})}\BibitemShut {NoStop}%
\bibitem [{\citenamefont {Shailos}\ \emph {et~al.}(2007)\citenamefont
  {Shailos}, \citenamefont {Nativel}, \citenamefont {Kasumov}, \citenamefont
  {Collet}, \citenamefont {Ferrier}, \citenamefont {Guéron}, \citenamefont
  {Deblock},\ and\ \citenamefont {Bouchiat}}]{Shailos_EPL_79_57008_2007}%
  \BibitemOpen
  \bibfield  {author} {\bibinfo {author} {\bibfnamefont {A.}~\bibnamefont
  {Shailos}}, \bibinfo {author} {\bibfnamefont {W.}~\bibnamefont {Nativel}},
  \bibinfo {author} {\bibfnamefont {A.}~\bibnamefont {Kasumov}}, \bibinfo
  {author} {\bibfnamefont {C.}~\bibnamefont {Collet}}, \bibinfo {author}
  {\bibfnamefont {M.}~\bibnamefont {Ferrier}}, \bibinfo {author} {\bibfnamefont
  {S.}~\bibnamefont {Guéron}}, \bibinfo {author} {\bibfnamefont
  {R.}~\bibnamefont {Deblock}}, \ and\ \bibinfo {author} {\bibfnamefont
  {H.}~\bibnamefont {Bouchiat}},\ }\href
  {http://stacks.iop.org/0295-5075/79/i=5/a=57008} {\bibfield  {journal}
  {\bibinfo  {journal} {EPL}\ }\textbf {\bibinfo {volume} {79}},\ \bibinfo
  {pages} {57008} (\bibinfo {year} {2007})}\BibitemShut {NoStop}%
\bibitem [{\citenamefont {Du}\ \emph {et~al.}(2008)\citenamefont {Du},
  \citenamefont {Skachko},\ and\ \citenamefont
  {Andrei}}]{Du_PRB_77_184507_2008}%
  \BibitemOpen
  \bibfield  {author} {\bibinfo {author} {\bibfnamefont {X.}~\bibnamefont
  {Du}}, \bibinfo {author} {\bibfnamefont {I.}~\bibnamefont {Skachko}}, \ and\
  \bibinfo {author} {\bibfnamefont {E.~Y.}\ \bibnamefont {Andrei}},\ }\href
  {\doibase 10.1103/PhysRevB.77.184507} {\bibfield  {journal} {\bibinfo
  {journal} {Phys. Rev. B}\ }\textbf {\bibinfo {volume} {77}},\ \bibinfo
  {pages} {184507} (\bibinfo {year} {2008})}\BibitemShut {NoStop}%
\bibitem [{\citenamefont {Mizuno}\ \emph {et~al.}(2013)\citenamefont {Mizuno},
  \citenamefont {Nielsen},\ and\ \citenamefont
  {Du}}]{Mizuno_Nat_Commun_4_2716_2013}%
  \BibitemOpen
  \bibfield  {author} {\bibinfo {author} {\bibfnamefont {N.}~\bibnamefont
  {Mizuno}}, \bibinfo {author} {\bibfnamefont {B.}~\bibnamefont {Nielsen}}, \
  and\ \bibinfo {author} {\bibfnamefont {X.}~\bibnamefont {Du}},\ }\href
  {http://dx.doi.org/10.1038/ncomms3716} {\bibfield  {journal} {\bibinfo
  {journal} {Nat. Commun.}\ }\textbf {\bibinfo {volume} {4}},\ \bibinfo {pages}
  {2716} (\bibinfo {year} {2013})}\BibitemShut {NoStop}%
\bibitem [{\citenamefont {Titov}\ and\ \citenamefont
  {Beenakker}(2006)}]{josephsongraphene1}%
  \BibitemOpen
  \bibfield  {author} {\bibinfo {author} {\bibfnamefont {M.}~\bibnamefont
  {Titov}}\ and\ \bibinfo {author} {\bibfnamefont {C.~W.~J.}\ \bibnamefont
  {Beenakker}},\ }\href {\doibase 10.1103/PhysRevB.74.041401} {\bibfield
  {journal} {\bibinfo  {journal} {Phys. Rev. B}\ }\textbf {\bibinfo {volume}
  {74}},\ \bibinfo {pages} {041401} (\bibinfo {year} {2006})}\BibitemShut
  {NoStop}%
\bibitem [{\citenamefont {Beenakker}(2008)}]{beenakker_RMP_80_1337_2008}%
  \BibitemOpen
  \bibfield  {author} {\bibinfo {author} {\bibfnamefont {C.~W.~J.}\
  \bibnamefont {Beenakker}},\ }\href {\doibase 10.1103/RevModPhys.80.1337}
  {\bibfield  {journal} {\bibinfo  {journal} {Rev. Mod. Phys.}\ }\textbf
  {\bibinfo {volume} {80}},\ \bibinfo {pages} {1337} (\bibinfo {year}
  {2008})}\BibitemShut {NoStop}%
\bibitem [{\citenamefont {Olund}\ and\ \citenamefont {Zhao}(2012)}]{Olund2012}%
  \BibitemOpen
  \bibfield  {author} {\bibinfo {author} {\bibfnamefont {C.~T.}\ \bibnamefont
  {Olund}}\ and\ \bibinfo {author} {\bibfnamefont {E.}~\bibnamefont {Zhao}},\
  }\href {\doibase 10.1103/PhysRevB.86.214515} {\bibfield  {journal} {\bibinfo
  {journal} {Phys. Rev. B}\ }\textbf {\bibinfo {volume} {86}},\ \bibinfo
  {pages} {214515} (\bibinfo {year} {2012})}\BibitemShut {NoStop}%
\bibitem [{\citenamefont {Ghaemi}\ and\ \citenamefont
  {Nair}(2016)}]{Ghaemi2016}%
  \BibitemOpen
  \bibfield  {author} {\bibinfo {author} {\bibfnamefont {P.}~\bibnamefont
  {Ghaemi}}\ and\ \bibinfo {author} {\bibfnamefont {V.~P.}\ \bibnamefont
  {Nair}},\ }\href {\doibase 10.1103/PhysRevLett.116.037001} {\bibfield
  {journal} {\bibinfo  {journal} {Phys. Rev. Lett.}\ }\textbf {\bibinfo
  {volume} {116}},\ \bibinfo {pages} {037001} (\bibinfo {year}
  {2016})}\BibitemShut {NoStop}%
\bibitem [{\citenamefont {Chen}\ \emph {et~al.}(2013)\citenamefont {Chen},
  \citenamefont {Jiang}, \citenamefont {Shen}, \citenamefont {Sheng},
  \citenamefont {Wang},\ and\ \citenamefont {Xing}}]{andreevweyl1}%
  \BibitemOpen
  \bibfield  {author} {\bibinfo {author} {\bibfnamefont {W.}~\bibnamefont
  {Chen}}, \bibinfo {author} {\bibfnamefont {L.}~\bibnamefont {Jiang}},
  \bibinfo {author} {\bibfnamefont {R.}~\bibnamefont {Shen}}, \bibinfo {author}
  {\bibfnamefont {L.}~\bibnamefont {Sheng}}, \bibinfo {author} {\bibfnamefont
  {B.~G.}\ \bibnamefont {Wang}}, \ and\ \bibinfo {author} {\bibfnamefont
  {D.~Y.}\ \bibnamefont {Xing}},\ }\href
  {http://stacks.iop.org/0295-5075/103/i=2/a=27006} {\bibfield  {journal}
  {\bibinfo  {journal} {EPL}\ }\textbf {\bibinfo {volume} {103}},\ \bibinfo
  {pages} {27006} (\bibinfo {year} {2013})}\BibitemShut {NoStop}%
\bibitem [{\citenamefont {Beenakker}(1991)}]{beenakker_PRL67_3836_1991}%
  \BibitemOpen
  \bibfield  {author} {\bibinfo {author} {\bibfnamefont {C.~W.~J.}\
  \bibnamefont {Beenakker}},\ }\href {\doibase 10.1103/PhysRevLett.67.3836}
  {\bibfield  {journal} {\bibinfo  {journal} {Phys. Rev. Lett.}\ }\textbf
  {\bibinfo {volume} {67}},\ \bibinfo {pages} {3836} (\bibinfo {year}
  {1991})}\BibitemShut {NoStop}%
\bibitem [{\citenamefont
  {Beenakker}(1992)}]{beenakker_in_Transport_Phenomena_in_Mesoscopic_Systems_eds_H_Fukuyama_ad_T_Ando_Springer_Berlin_1992}%
  \BibitemOpen
  \bibfield  {author} {\bibinfo {author} {\bibfnamefont {C.~W.~J.}\
  \bibnamefont {Beenakker}},\ }in\ \href {\doibase
  10.1007/978-3-642-84818-6_22} {\emph {\bibinfo {booktitle} {Transport
  Phenomena in Mesoscopic Systems}}},\ \bibinfo {editor} {edited by\ \bibinfo
  {editor} {\bibfnamefont {H.}~\bibnamefont {Fukuyama}}\ and\ \bibinfo {editor}
  {\bibfnamefont {T.}~\bibnamefont {Ando}}}\ (\bibinfo  {publisher} {Springer,
  Berlin/Heidelberg},\ \bibinfo {year} {1992})\ pp.\ \bibinfo {pages}
  {235--253}\BibitemShut {NoStop}%
\bibitem [{\citenamefont {Brouwer}\ and\ \citenamefont
  {Beenakker}(1997)}]{josephsontemperature}%
  \BibitemOpen
  \bibfield  {author} {\bibinfo {author} {\bibfnamefont {P.~W.}\ \bibnamefont
  {Brouwer}}\ and\ \bibinfo {author} {\bibfnamefont {C.~W.~J.}\ \bibnamefont
  {Beenakker}},\ }\href {\doibase
  http://dx.doi.org/10.1016/S0960-0779(97)00018-0} {\bibfield  {journal}
  {\bibinfo  {journal} {Chaos Soliton Fract}\ }\textbf {\bibinfo {volume}
  {8}},\ \bibinfo {pages} {1249 } (\bibinfo {year} {1997})}\BibitemShut
  {NoStop}%
\bibitem [{\citenamefont {{Tworzyd{\l}o}}\ \emph {et~al.}(2006)\citenamefont
  {{Tworzyd{\l}o}}, \citenamefont {{Trauzettel}}, \citenamefont {{Titov}},
  \citenamefont {{Rycerz}},\ and\ \citenamefont {{Beenakker}}}]{rngraphene}%
  \BibitemOpen
  \bibfield  {author} {\bibinfo {author} {\bibfnamefont {J.}~\bibnamefont
  {{Tworzyd{\l}o}}}, \bibinfo {author} {\bibfnamefont {B.}~\bibnamefont
  {{Trauzettel}}}, \bibinfo {author} {\bibfnamefont {M.}~\bibnamefont
  {{Titov}}}, \bibinfo {author} {\bibfnamefont {A.}~\bibnamefont {{Rycerz}}}, \
  and\ \bibinfo {author} {\bibfnamefont {C.~W.~J.}\ \bibnamefont
  {{Beenakker}}},\ }\href {\doibase 10.1103/PhysRevLett.96.246802} {\bibfield
  {journal} {\bibinfo  {journal} {Phys. Rev. Lett.}\ }\textbf {\bibinfo
  {volume} {96}},\ \bibinfo {eid} {246802} (\bibinfo {year}
  {2006})}\BibitemShut {NoStop}%
\bibitem [{\citenamefont {Baireuther}\ \emph {et~al.}(2014)\citenamefont
  {Baireuther}, \citenamefont {Edge}, \citenamefont {Fulga}, \citenamefont
  {Beenakker},\ and\ \citenamefont {Tworzyd\l{}o}}]{Baireuther_PRB_89_035410}%
  \BibitemOpen
  \bibfield  {author} {\bibinfo {author} {\bibfnamefont {P.}~\bibnamefont
  {Baireuther}}, \bibinfo {author} {\bibfnamefont {J.~M.}\ \bibnamefont
  {Edge}}, \bibinfo {author} {\bibfnamefont {I.~C.}\ \bibnamefont {Fulga}},
  \bibinfo {author} {\bibfnamefont {C.~W.~J.}\ \bibnamefont {Beenakker}}, \
  and\ \bibinfo {author} {\bibfnamefont {J.}~\bibnamefont {Tworzyd\l{}o}},\
  }\href {\doibase 10.1103/PhysRevB.89.035410} {\bibfield  {journal} {\bibinfo
  {journal} {Phys. Rev. B}\ }\textbf {\bibinfo {volume} {89}},\ \bibinfo
  {pages} {035410} (\bibinfo {year} {2014})}\BibitemShut {NoStop}%
\bibitem [{\citenamefont {Sbierski}\ \emph {et~al.}(2014)\citenamefont
  {Sbierski}, \citenamefont {Pohl}, \citenamefont {Bergholtz},\ and\
  \citenamefont {Brouwer}}]{Sbierski_PRL_113_026602}%
  \BibitemOpen
  \bibfield  {author} {\bibinfo {author} {\bibfnamefont {B.}~\bibnamefont
  {Sbierski}}, \bibinfo {author} {\bibfnamefont {G.}~\bibnamefont {Pohl}},
  \bibinfo {author} {\bibfnamefont {E.~J.}\ \bibnamefont {Bergholtz}}, \ and\
  \bibinfo {author} {\bibfnamefont {P.~W.}\ \bibnamefont {Brouwer}},\ }\href
  {\doibase 10.1103/PhysRevLett.113.026602} {\bibfield  {journal} {\bibinfo
  {journal} {Phys. Rev. Lett.}\ }\textbf {\bibinfo {volume} {113}},\ \bibinfo
  {pages} {026602} (\bibinfo {year} {2014})}\BibitemShut {NoStop}%
\bibitem [{\citenamefont {Bergholtz}\ \emph {et~al.}(2015)\citenamefont
  {Bergholtz}, \citenamefont {Liu}, \citenamefont {Trescher}, \citenamefont
  {Moessner},\ and\ \citenamefont {Udagawa}}]{Bergholtz2015}%
  \BibitemOpen
  \bibfield  {author} {\bibinfo {author} {\bibfnamefont {E.~J.}\ \bibnamefont
  {Bergholtz}}, \bibinfo {author} {\bibfnamefont {Z.}~\bibnamefont {Liu}},
  \bibinfo {author} {\bibfnamefont {M.}~\bibnamefont {Trescher}}, \bibinfo
  {author} {\bibfnamefont {R.}~\bibnamefont {Moessner}}, \ and\ \bibinfo
  {author} {\bibfnamefont {M.}~\bibnamefont {Udagawa}},\ }\href {\doibase
  10.1103/PhysRevLett.114.016806} {\bibfield  {journal} {\bibinfo  {journal}
  {Phys. Rev. Lett.}\ }\textbf {\bibinfo {volume} {114}},\ \bibinfo {pages}
  {016806} (\bibinfo {year} {2015})}\BibitemShut {NoStop}%
\bibitem [{\citenamefont {Trescher}\ \emph {et~al.}(2015)\citenamefont
  {Trescher}, \citenamefont {Sbierski}, \citenamefont {Brouwer},\ and\
  \citenamefont {Bergholtz}}]{Trescher2015}%
  \BibitemOpen
  \bibfield  {author} {\bibinfo {author} {\bibfnamefont {M.}~\bibnamefont
  {Trescher}}, \bibinfo {author} {\bibfnamefont {B.}~\bibnamefont {Sbierski}},
  \bibinfo {author} {\bibfnamefont {P.~W.}\ \bibnamefont {Brouwer}}, \ and\
  \bibinfo {author} {\bibfnamefont {E.~J.}\ \bibnamefont {Bergholtz}},\ }\href
  {\doibase 10.1103/PhysRevB.91.115135} {\bibfield  {journal} {\bibinfo
  {journal} {Phys. Rev. B}\ }\textbf {\bibinfo {volume} {91}},\ \bibinfo
  {pages} {115135} (\bibinfo {year} {2015})}\BibitemShut {NoStop}%
\bibitem [{\citenamefont {{Soluyanov}}\ \emph {et~al.}(2015)\citenamefont
  {{Soluyanov}}, \citenamefont {{Gresch}}, \citenamefont {{Wang}},
  \citenamefont {{Wu}}, \citenamefont {{Troyer}}, \citenamefont {{Dai}},\ and\
  \citenamefont {{Bernevig}}}]{Soluyanov2015}%
  \BibitemOpen
  \bibfield  {author} {\bibinfo {author} {\bibfnamefont {A.~A.}\ \bibnamefont
  {{Soluyanov}}}, \bibinfo {author} {\bibfnamefont {D.}~\bibnamefont
  {{Gresch}}}, \bibinfo {author} {\bibfnamefont {Z.}~\bibnamefont {{Wang}}},
  \bibinfo {author} {\bibfnamefont {Q.}~\bibnamefont {{Wu}}}, \bibinfo {author}
  {\bibfnamefont {M.}~\bibnamefont {{Troyer}}}, \bibinfo {author}
  {\bibfnamefont {X.}~\bibnamefont {{Dai}}}, \ and\ \bibinfo {author}
  {\bibfnamefont {B.~A.}\ \bibnamefont {{Bernevig}}},\ }\href {\doibase
  10.1038/nature15768} {\bibfield  {journal} {\bibinfo  {journal} {Nature
  (London)}\ }\textbf {\bibinfo {volume} {527}},\ \bibinfo {pages} {495}
  (\bibinfo {year} {2015})}\BibitemShut {NoStop}%
\end{thebibliography}%

\end{document}